\documentclass[twocolumn,showpacs,preprintnumbers,amsmath,amssymb,natbib,noshowpacs]{revtex4}

\usepackage{graphicx}
\usepackage{dcolumn}
\usepackage{bm}
\usepackage{color}
\usepackage[colorlinks=true]{hyperref}
\usepackage{amssymb}
\usepackage{amsmath,empheq}
\usepackage{subfigure}

\newcommand{\be}{\begin{equation}}
\newcommand{\ee}{\end{equation}}
\newcommand{\bear}{\begin{eqnarray}}
\newcommand{\eear}{\end{eqnarray}}

\newcommand{\ba}{\begin{array}}
\newcommand{\ea}{\end{array}}

\begin{document}
\preprint{OU-HET-964}

\title{
Chaos of Wilson Loop \\
from String Motion near Black Hole Horizon}

\author{Koji Hashimoto}
\author{Keiju Murata}
\affiliation{Department of Physics, Osaka University, Toyonaka, Osaka 560-0043, Japan}
\author{Norihiro Tanahashi}
\affiliation{Institute of Mathematics for Industry, University of Kyushu, 744 Motooka, Nishi-ku, Fukuoka 819-0395, Japan}


\begin{abstract} 
To find the origin of chaos near black hole horizon in string-theoretic AdS/CFT correspondence,
we perform a chaos analysis of a suspended string in AdS black hole backgrounds.
It has a definite CFT interpretation: chaos of Wilson loops, or in other words, sensitive time-evolution
of a quark antiquark force in thermal gauge theories. 
Our nonlinear numerical simulation of the suspended Nambu-Goto string shows chaos,
which would be absent in pure AdS background.
The calculated Lyapunov exponent $\lambda$ satisfies
the universal bound $\lambda \leq 2\pi T_{\rm H}$ for the Hawking temperature $T_{\rm H}$.
We also analyze a toy model of a rectangular string probing the horizon and show that it contains
a universal saddle characterized by the surface gravity $2\pi T_{\rm H}$.
Our work demonstrates that the black hole horizon is the origin of the chaos,
and suggests a close interplay between chaos and quark deconfinement.
\end{abstract}


\maketitle

\setcounter{footnote}{0}



\section{Introduction}
\label{Sec:intro}

The renowned AdS/CFT correspondence \cite{Maldacena:1997re}
asserts that a certain class of quantum field theories 
allows a corresponding gravity description. 
Since the classical gravity is characterized by its nonlinear solutions: black holes,
characteristics of black hole horizons are believed to be a key to
unveil the unknown mechanism of how the AdS/CFT works.
In the course of research on 
shock waves near black hole horizons \cite{Shenker:2013pqa,Shenker:2013yza}
(and subsequent study \cite{Leichenauer:2014nxa,Kitaev-talk,Shenker:2014cwa,Jackson:2014nla,Polchinski:2015cea}),
Maldacena, Shenker and Stanford
provided \cite{Maldacena:2015waa} a universal chaos bound for 
the Lyapunov exponent $\lambda$ of 
out-of-time-ordered correlators \cite{Larkin,Shenker:2013pqa} in thermal quantum field theories,
\begin{align}
\lambda \leq \frac{2\pi T}{\hbar} \, .
\label{MSS}
\end{align}
Here $T$ is the temperature, and in the gravity dual it is the Hawking temperature of the black hole,
equal to the surface gravity at the horizon.

The issue here is how the chaos of a fundamental string is derived near black hole horizons.
The AdS/CFT correspondence is best understood in string theory, and the fundamental object
is string theory is of course the fundamental string. 
In this paper we provide a concrete chaos analysis
of motion of a fundamental string very near the black hole horizon.

Although chaos in string motion has been studied 
\cite{Basu:2011dg,T11,WQCD,D-brane,complex-beta,NR,T11-ppwave,gamma,BH,AKY,Ishii:2016rlk} 
in curved geometries, in view of the
chaos bound \eqref{MSS}, two important issues are
left unsolved: First, the motion of the string needs to be near the black hole horizon, and second,
the CFT interpretation of the string motion is indispensable. In this paper we address these two
issues, by adopting a suspended string in the AdS black hole geometry. 

One of the advantage to consider the suspended string 
is that it can probe near the black hole horizon in a natural way.
In \cite{Hashimoto:2016dfz}, a particle probing the region near the black hole horizon, pulled outward
by some external force, was shown to have a universal behavior around the local maximum of the
energy. 
The suspended string provides a natural set-up for the universal chaos of the black hole horizon in AdS/CFT, 
by replacing the particle with a Nambu-Goto string, a natural object
in the top-down AdS/CFT correspondence.
In Refs.~\cite{Ishii:2015wua,Ishii:2015qmj}, 
dynamics of Nambu-Goto string in pure AdS background has been studied. 
For pure AdS, there was no sensitivity to initial conditions  
although the energy cascade in the string worldsheet was observed.
We expect that the effect of the event horizon to chaos 
can be extracted by the analysis of the suspended string dynamics in the black hole spacetime.

The other advantage is that the suspended  string has a clear CFT 
interpretation~\cite{Maldacena:1998im,Rey:1998ik,Rey:1998bq}: a Wilson loop in thermal Yang-Mills theory,
in other words, an interquark force.
The dual CFT observable we calculate is the Lyapunov exponent $\lambda$ 
of chaos of the interquark force $\vec{F}$: 
\begin{align}
\langle \delta \vec{F} (t) \rangle \simeq e^{\lambda t}\langle \delta \vec{F} (0) \rangle 
\label{deltaF12}
\end{align}
where $\langle \delta \vec{F}(t)\rangle \equiv \langle \vec{F}(t)\rangle_1-\langle \vec{F}(t)\rangle_2$ is 
the difference of the interquark forces due to two different initial conditions indicated by 
$\langle \rangle_1$ and $\langle \rangle_2$ (the initial ($t\sim 0$) infinitesimal kicks about the
interquark distance $L(t)$ are chosen to be slightly different).
In the time evolution, the difference grows exponentially for chaotic systems, with the
Lyapunov exponent $\lambda >0$. 

Chaos of Wilson loops is important for understanding meson dynamics and deconfinement transition.
The charmonium suppression in heavy ion collisions may be understood by an entropic interquark 
force \cite{Kharzeev:2014pha,Hashimoto:2014fha,Iatrakis:2015sua,Fadafan:2015ynz}, 
and the chaos could provide the entropy
in a manner similar to \cite{Iida:2013qwa,Tsukiji:2016krj,Tsukiji:2017pjx}.
At the deconfinement a turbulent behavior is expected \cite{Fukushima:2013dma,Fukushima:2016xgg}, 
and the scaling of excited meson numbers 
\cite{Hashimoto:2014xta,Hashimoto:2014dda,Hashimoto:2015psa} 
can be attributed to the chaos of mesons \cite{Hashimoto:2016wme} and chaos of the QCD string.

Let us briefly review the suspended string in AdS black hole geometry, with emphasis 
also on an unstable solution which probes the horizon.
Consider a fundamental string suspended from the boundary of the AdS black hole geometry.
The background metric in 10-dimensional spacetime is 
\begin{align}
ds^2 = \frac{r^2}{R^2}
\left[-f(r) dt^2 + d\vec{x}^2
\right] + \frac{R^2}{r^2f(r)} dr^2 + R^2 d\Omega_5^2
\label{metric}
\end{align}
with $f(r)\equiv 1-r_H^4/r^4$, and $R$ is the AdS radius. The last term is for the internal $S^5$
for the geometry to be a background of the type IIB superstring theory.
The horizon is located at $r=r_H$, and the temperature of the black hole is given as
$T_H = r_H/\pi R^2$.
The action for a fundamental string in this AdS black hole geometry is simply given by
\begin{align}
S = -{\cal T}\int\! d\tau d\sigma \sqrt{-\det(G_{\mu\nu}[X]\partial_\alpha X^\mu \partial_\beta X^\nu)}.
\label{acF}
\end{align}
Here $\mu,\nu = 0,1,2,3,r$ is the target space index and $\alpha,\beta=\tau,\sigma$
is the worldsheet index. We study the dynamics of the fundamental string in this geometry,
in particular when a part of the string is close to the black hole horizon $r=r_H$.

\begin{figure}[t]
	\begin{center}
\includegraphics[scale=0.66]{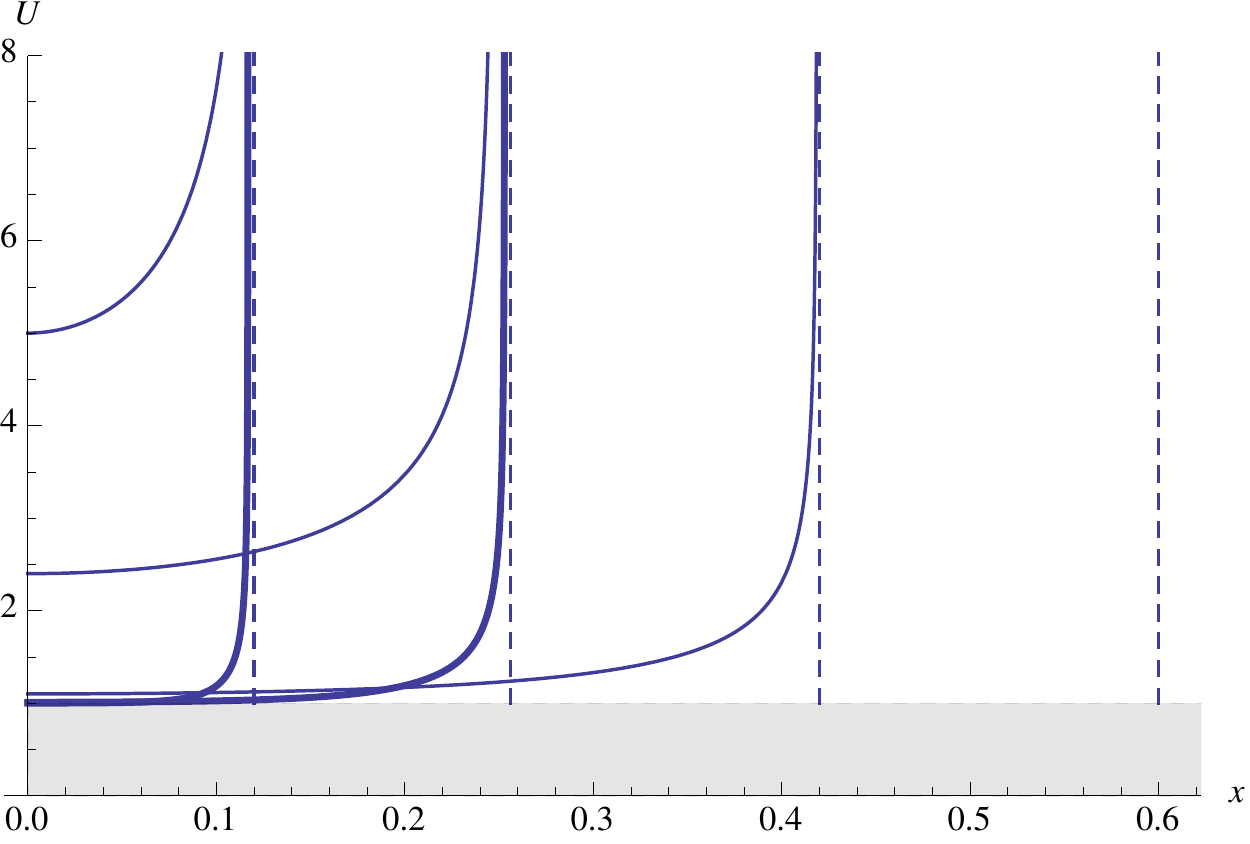}
	 \caption{The shape of a string suspended down to the near-horizon region. We show a right half of the strings,
	 for four choices of the asymptotic interquark separation $L$.
	 Thin solid lines are connected strings at the local minima of the energy, 
	 and thick solid lines are connected strings at the local maxima of the energy. Dashed lines are strings stick to
	 the black hole horizon. The shaded region is the black hole, and the line $U\equiv r/r_H=1$ is the horizon.}
	 \label{figshape}
          \end{center}
\label{Fig:static}
\end{figure}

In the AdS/CFT correspondence, 
a string suspended from the boundary $r=\infty$ of the geometry 
corresponds to
a pair of a quark and an anti-quark separated by the distance $L$ in the 4-dimensional CFT.
The energy
of the string is a quark-antiquark potential of ${\cal N}=4$ super Yang-Mills
theory.

In Fig.~\ref{figshape}, we show three possible shapes of the string for a fixed interquark separation $L$.
The figure describes one half part of the strings, since the shape is symmetric
under a parity exchanging the quark and the antiquark.
When $L$ is small (below some critical value), there are three possible configurations
of the string: solid line, solid thick line and dashed line in the figure. 
The dashed line is just a set of two parallel straight strings aligned along the $r$ direction, 
separated by the distance $L$, stick to the black hole horizon. Solid lines are a suspended string whose
asymptotic location is separated by the distance $L$. In particular, the solid non-thick line 
describes a local minimum of the
energy, while the thick one is at a local maximum of the energy, for a given fixed $L$. 
When $L$ is large, the configuration with the solid thick line does not exist.
The lowest energy configuration is the pair of the dashed strings written as the dashed line,
and the interquark force disappears. 

The goal of this paper is to calculate a Lyapunov exponent of 
a full numerical simulation of the suspended Nambu-Goto string.
The suspended string has a clear interpretation in the CFT side: a Wilson loop.
So the chaos described by the string is quantum chaos probed by the Wilson loop operator of 
non-Abelian gauge theories. In other words, 
the time evolution of the force between the quarks is exponentially sensitive to the 
difference of the initial conditions.
Our calculation is a prediction for the time evolution of the interquark forces at finite temperature in ${\cal N}=4$
supersymmetric Yang-Mills theory at large $N$ and strong coupling limit.

To study the origin of the chaos, we look at
the solid thick line of the string in Fig.~\ref{Fig:static} since it probes very near the black hole horizon.
This configuration has not been studied in literature simply because it is not the minimum energy configuration.
The unstable classical configuration of the string at the local maximum of the energy, is provided by
the balance of the pulling force
outward by the string tension and the gravity by the black hole, so it provides a natural realization of
the particle model given in \cite{Hashimoto:2016dfz}.

We provide three kinds of chaos analyses in this paper, and find a chaos nested near the black hole horizon.
First, in Section \ref{sec:universal},
we provide a toy model in which the shape of the suspended string is of a rectangular shape. This 
approximation is natural in view of the true shape of the solid thick lines in Fig.~\ref{figshape}.
The toy model shows the existence of the local maximum of the energy, at which the 
Lyapunov exponent of the chaos is evaluated to be universal and saturates the 
bound $\lambda = 2\pi T_{\rm H}$.
In Appendix~\ref{App:pq-string} we show that
a more generic $(p,q)$-string in general near-horizon geometry of
D3-F1-D1 bound states shows the same Lyapunov exponent, 
thus this value is universal.

Next, in Section \ref{sec:pert}, we provide a chaos analysis of the fluctuation modes of the string. When the
minimum energy configuration of the suspended string (the non-thick line in Fig.~\ref{figshape})
fluctuates, the motion is expected to show chaos. We truncate the fluctuation to just the lowest and the
first excited states, and introduce cubic order interaction calculated from the Nambu-Goto action; it is
the minimal set of modes and interaction to show chaos. 
We find numerical evidence that chaos emerges due to the effect of the black hole horizon even in this simplified set-up.

Finally, in Section \ref{sec:nonlinear}, we provide a full nonlinear study of the Nambu-Goto string in the black hole
background. With a ``kick'' at the boundary of the stable suspended string, the string starts to move in the curved geometry. We look at the time-dependent force which the quark feels, and also the 
motion of the bottom of the suspended string, and extract the Lyapunov exponent.
We find that the measured 
Lyapunov exponent satisfies the universal bound, $\lambda \sim 0.065\times 2\pi T_{\rm H}$.
This is our prediction for the chaos of the interquark forces in thermal gauge theories with an AdS-Schwarzschild 
gravity dual.

The last section is devoted for a summary and discussions on the interpretation of our Lyapunov exponents
as a CFT out-of-time-order correlator.

\section{Universality in a toy string model}
\label{sec:universal}

In this section, we present a string toy model which shows a universal chaos.
The motion of the string is quite complicated since it 
is not integrable and is subject to a set of partial differential equations following from
the action \eqref{acF}.
To find a universal feature of the dynamics near the horizon, it is instructive to start with
an approximation for the shape of the string. In view of Fig.~\ref{figshape}, 
we can naturally  approximate the configuration of the string by a rectangular shape.

Suppose that the bottom part of the rectangle is at $r=r_0$, then the total action 
of the string is given by
\begin{align}
S & = -{\cal T}\int dt \left[\sqrt{g_{00}(r_0) g_{xx}(r_0)} L - 2\int_{r_H}^{r_0}
\sqrt{g_{00}(r) g_{rr}(r)} dr
\right] \nonumber \\
& 
= -{\cal T}\int dt \left[\frac{L}{R^2} \sqrt{r_0^4-r_H^4}  - 2 (r_0-r_H)
\right] \, .
\end{align}
The first term is for the bottom part of the string, and the second term is
the contribution from the vertical parts (the parallel two edges connecting
$r=r_0$ and the AdS boundary). Note that the second term is the remaining
after subtracting the infinite contribution $2\int_{r_0}^\infty dr$ which is the
length of the infinitely long strings, corresponding to the action of the two independent 
quarks.

The total energy of the string is then
\begin{align}
{\cal E}
= {\cal T} \left[\frac{L}{R^2} \sqrt{r_0^4-r_H^4}  - 2 (r_0-r_H) \right]
\, .
\end{align}
Extremizing this energy means
\begin{align}
r_0^6 = \frac{R^4}{L^2}(r_0^4-r_H^4) \, .
\end{align}
We are interested in the case $r_0$ approaching the horizon, and in fact for small $L$
this equation has a solution near the horizon.
For small $L$, the solution is expanded as
\begin{align}
r_0^{\rm sol} = r_H 
\left(
1 + \frac14 \left(\frac{r_H}{R^2} L\right)^2
+ \frac9{32} \left(\frac{r_H}{R^2} L\right)^4 + \cdots
\right) \, .
\label{r0sol}
\end{align}
This solution actually maximizes the energy locally, since if we expand the energy
for small change of the location of the bottom of the string,
\begin{align}
r_0 = r_0^{\rm sol} + \delta r \, ,
\end{align}
then
\begin{align}
{\cal E} = \mbox{const.} - {\cal T} \frac{2R^4}{r_H^3 L^2} (\delta r)^2 + \cdots \, .
\label{enest}
\end{align}
So, the solution with $r_0 = r_0^{\rm sol}$ of \eqref{r0sol} is at the top of the potential hill,
maximizing the energy locally.

\begin{figure}[t]
	\begin{center}
\includegraphics[scale=0.66]{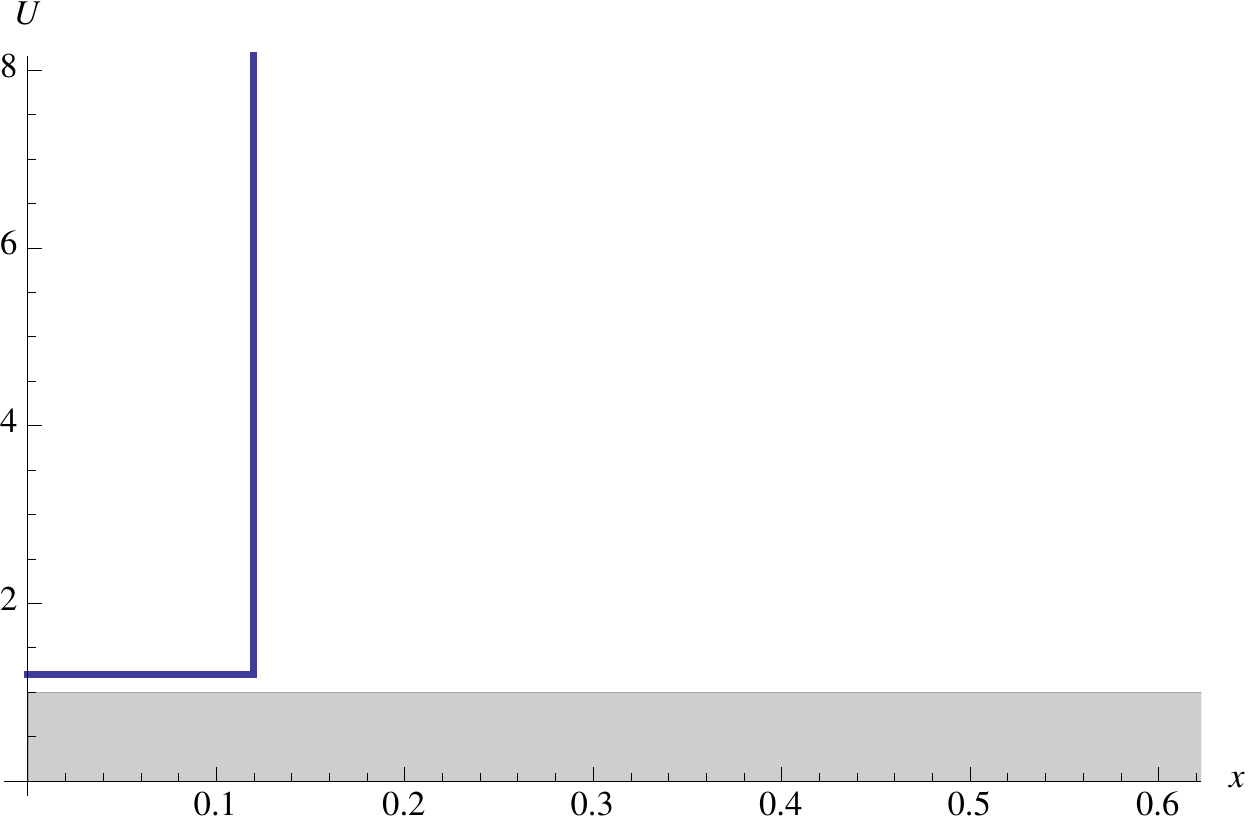}
	 \caption{Our approximate configuration of the string of the rectangular shape.}
	 \label{figshape2}
          \end{center}
\label{Fig:rect}
\end{figure}

To look at the motion of the string, we evaluate the kinetic energy of
the bottom part of the string moving slowly in the $r$ direction. When there
is a time dependence $r_0(t)$, the relevant part of the action is
\begin{align}
S = -{\cal T}\int dt \; L \sqrt{-(g_{00} + g_{rr}\dot{r}_0^2) g_{xx}} \, ,
\end{align}
which is expanded in terms of small $\dot{r}_0$ as
\begin{align}
S = -{\cal T}L \int \! dt \; \frac{r_0^2}{R^2}\sqrt{f(r_0)} \left(
1-\frac{R^4}{2r_0^4 f(r_0)^2}\dot{r}_0^2 + \cdots
\right) \, .
\end{align}
Evaluating the coefficient of the kinetic term $\dot{r}_0^2$ at $r_0^{\rm sol}$ for small $L$,
we obtain an effective action for the motion of the string, together with \eqref{enest},
\begin{align}
S = {\cal T} \frac{R^8}{2r_H^5 L^2} \left(\frac{d}{dt}\delta r\right)^2 + {\cal T} \frac{R^4}{r_H^3 L^2} (\delta r)^2 \, .
\end{align}
Using the relation to the Hawking temperature of the black hole, $r_H = \pi R^2 T_{\rm H}$, this effective action can be
simply written as
\begin{align}
S = {\cal T} \frac{R^8}{2r_H^5 L^2} 
\left[
\left(\frac{d}{dt}\delta r\right)^2 + (2\pi T_{\rm H})^2 (\delta r)^2
\right] \, .
\label{effst}
\end{align}
Here, the factor $2\pi T_{\rm H}$ is the surface gravity of the black hole.
A generic 
solution of the equation of motion derived from the effective action \eqref{effst} is given as
\begin{align}
\delta r = A \exp [2\pi T_{\rm H} t] + B \exp[-2\pi T_{\rm H} t]
\end{align}
whose first term is exponentially growing. Interpreting the maximum point of the potential 
as a separatrix as in \cite{Hashimoto:2016dfz}, we find that the Lyapunov exponent $\lambda$
is given by
\begin{align}
\lambda = 2\pi T_{\rm H} \, .
\end{align}
Therefore, the Lyapunov exponent is given by the Hawking temperature of the black hole.

Although the system depends on various parameters such as the string tension ${\cal T}$, 
the AdS radius $R$, the black hole radius
$r_H$ and the quark separation $L$, the Lyapunov exponent depends only on the Hawking temperature.
So the Lyapunov exponent appears to be universally characterized by the Hawking temperature, 
independent of the details of the system. In addition, the expression of the Lyapunov exponent saturates
the bound conjectured in \cite{Maldacena:2015waa}.

In Appendix~\ref{App:pq-string}, we find that the universality of the Lyapunov exponent persists even when the fundamental string is generalized to the $(p,q)$-string in the near-horizon geometry of D3-F1-D1 bound state at a finite temperature.

\section{Chaos of perturbative string motion}
\label{sec:pert}

In the previous section, we argued that the Lyapunov exponent for a string moving in the near-horizon region takes a universal value that depends only on the horizon temperature.
To make a firmer connection between this Lyapunov exponent and the chaotic behavior of the string, in the following sections we study the string motion and its chaotic behavior in more direct ways.
In this section, we study an effective model for the perturbative string motion by truncating up to the first excited state.
We also examine the full nonlinear dynamics of the string and evaluate the Lyapunov exponent in section~\ref{sec:nonlinear}.

To describe the perturbative string motion, we follow the method of Ref.~\cite{Arias:2009me}.
Introducing dimensionless coordinates by
$\bar t = (r_H/R^2) t,T$
$\bar r = r / r_H$,
$\bar x = (r_h/R^2) x$, 
the metric~(\ref{metric}) is expressed as
\begin{gather}
ds^2 = - g_{t}(\bar r)d\bar t^2 + g_r(\bar r) d \bar r^2 + g_x(\bar r) d \bar x^2,
\label{metric2}
\\
g_t(\bar r) = R^2(\bar r^2 -\bar r^{-2}),
~
g_r(\bar r) = \frac{R^2}{\bar r^2 - \bar r^{-2}},
~
g_x (\bar r) = R^2 \bar r^2.
\end{gather}
For simplicity we omit the bar of the dimensionless coordinates henceforth.

We first study a static string configuration. 
Specifying the string location by $(r,x) = (r(\ell), x(\ell))$, where $\ell$ is the proper distance measured along the string from its tip, the action for the string becomes
\begin{equation}
S = -R^2 {\cal T}\int d\ell 
\sqrt{
g^2(r) r'^2(\ell) + f^2(r) x'^2(\ell)
},
\label{action_static}
\end{equation}
where we defined
\begin{equation}
\begin{aligned}
f^2(r) &= g_t(r) g_x(r) = r^4-1,
\\
g^2(r) &= g_t(r) g_r(r) = 1,
\\
h^2(r) &= g_r(r) g_x(r) = \frac{r^4}{r^4-1}.
\end{aligned}
\end{equation}
Since $\ell$ is the proper length along the string, 
the tangent vector along the string $t^a = \left(r'(\ell),  x'(\ell)\right)$ becomes a unit vector, that is,
\begin{equation}
g_r(r) r'^2(\ell) + g_x(r) x'^2(\ell) = 1.
\label{eqproper}
\end{equation}
Also, the action~(\ref{action_static}) implies an equation of motion for a static string given by
\begin{equation}
\frac{dx}{dr} = 
\pm\frac{g(r)}{f(r)}
\frac{f(r_0)}{\sqrt{f^2(r)-f^2(r_0)}},
\label{geodeq}
\end{equation}
where $r = r_0$ is the location of the tip of the string.
Equations (\ref{eqproper}) and (\ref{geodeq}) fix the location of the static string $(r,x) = (r_\text{BG}(\ell),x_\text{BG}(\ell))$, which will be used as the background solution to study small fluctuations around it.
Below, we focus on the $x>0$ part of the string and choose the $+$ sign in Eq.~(\ref{geodeq}).

As explained in section~\ref{Sec:intro}, there are one stable and one unstable static configurations for the same boundary condition (the quark separation $L$). As the background solution used in this section, we take the unstable static string, which is constructed by setting $r_0$ sufficiently close to $r_H$.
Below, we consider perturbative motion around this unstable maximum taking into the next-to-leading order perturbations. 
The leading-order action describes an oscillator with an unstable potential, and the next-to-leading order perturbation generates a trapping potential nearby the unstable equilibrium point.
We consider the string motion in this trapping potential in the following analysis.

\begin{figure}[htb]
\centering
\includegraphics[width=5cm]{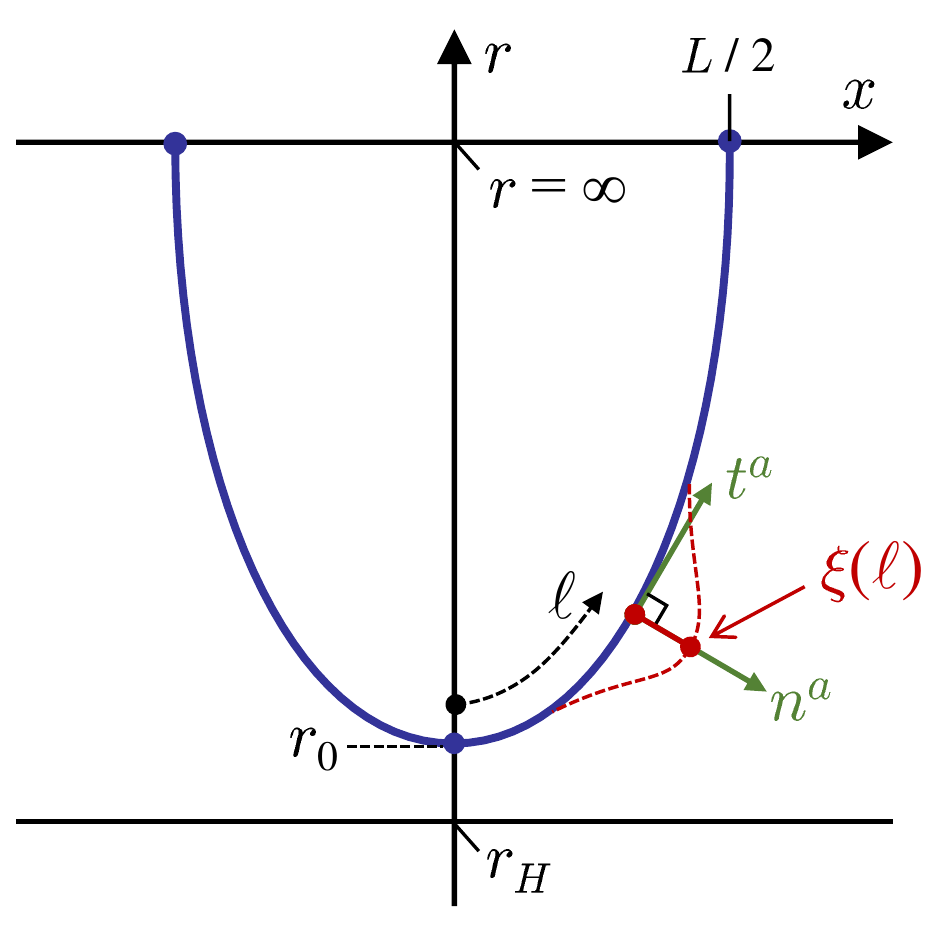}
\caption{Geometry of the perturbations around a static string.}
\label{Fig:n-gauge}
\end{figure}

To describe perturbations around the static string, we introduce the unit normal on the string by
\begin{equation}
n_a = h\bigl(r_\text{BG}(\ell)\bigr) \left(-x_\text{BG}'(\ell), r_\text{BG}'(\ell)\right).
\label{normal}
\end{equation}
We perturb the string toward the the normal direction $n^a$ by a proper distance $\xi(t,\ell)$. 
Then the target space coordinate of the perturbed string is given by (see Fig.~\ref{Fig:n-gauge})
\begin{equation}
r = r_\text{BG}(\ell) + \xi(t,\ell) n^r(\ell),
\quad
x = x_\text{BG}(\ell) + \xi(t,\ell) n^x(\ell).
\label{deformation}
\end{equation}

Expanding the action~(\ref{acF}) up to the third order in $\xi(t,\ell)$, we can construct an effective action for the perturbative string motion.
We defer the derivation to Appendix~\ref{App:Seff} and summarize only the construction procedure and the final result below.
The perturbation equation obtained from the quadratic action is given by 
\begin{equation}
\omega^2 C_{tt}(\ell) \xi(\ell) + 
\partial_\ell\left[ C_{\ell\ell}(\ell) \xi'(\ell)\right]
- C_{00}(\ell) \xi(\ell) = 0,
\label{xieq}
\end{equation}
where 
the coefficients $C_{tt}, C_{\ell\ell}, C_{00}$ are given by Eq.~(\ref{coeffs}), and 
we separated the variable as $\xi(t,\ell) = \xi(\ell) e^{i\omega t}$.
Solutions to this equation can be expanded using the eigenfunctions.
In our analysis, we truncate the solution up to the the first two excitations as
\begin{equation}
\xi(t,\ell) = c_0(t) e_0(\ell) + c_1(t) e_1(\ell),
\label{xi-ansatz(main)}
\end{equation}
where $e_{0,1}(\ell)$ are the normalized mode functions with the lowest and the next lowest eigenfrequencies.
Next, we evaluate the cubic-order part of the action using the truncated solution~(\ref{xi-ansatz(main)}), which results in
\begin{align}
&\frac{S^{(3)}}{\cal T} =
\int dt
\int_{-\infty}^\infty d\ell \Biggl\{
\left[\left( C_0 - \frac{C_1'}{3} \right) e_0^3 + C_2 e_0 e_0'\right] c_0^3(t)
\notag \\
&+ \left[
3 \left( C_0 - \frac{C_1'}{3} \right) e_0 e_1^2
+ C_2 \left(
e_0 e_1'^2 + 2 e_0' e_1 e_1'
\right)
\right]c_0(t)c_1^2(t)
\notag \\
&+ C_3e_0^3 c_0 \dot c_0^2
+ C_3 e_0 e_1^2 c_0 \dot c_1^2 
+ 2 C_3 e_0 e_1^2 \dot c_0 c_1 \dot c_1
\Biggr\},
\label{Scubic}
\end{align}
where $C_{0,1,2,3}(\ell)$ are functions of the background solution $r_\text{BG}(\ell)$.
Evaluating the action~(\ref{Scubic}) numerically we can construct the cubic action.
For example, when the tip of the string is at $r=r_0 = 1.1$, the total perturbative action is given by
\begin{multline}
\frac{S}{\cal T} = \int dt
\sum_{n=0,1}
\left(
 \dot{\tilde c}_n^2 - \omega_n^2 {\tilde c}_n^2
\right)
+ 7.11 \tilde c_0^3
+ 35.3 \tilde c_0 \tilde c_1^2 
\\
+ 4.66 \tilde c_0 \dot {\tilde c}_0^2 
+ 1.32 \tilde c_0 \dot {\tilde c}_1^2 
- 7.57 \dot {\tilde c}_0 \tilde c_1 \dot {\tilde c}_1,
\label{S_stabilized_deep}
\end{multline}
where $\omega_0^2 = - 1.40$ and $\omega_1^2 = 7.57$,
and $\tilde c_{0,1}(t)$ are functions of $c_{0,1}(t)$ introduced by Eq.~(\ref{tildecdef}) to stabilize the time evolution.

\begin{figure}[htb!]
\subfigure[Poincar\'e plot]{
\includegraphics[width=8cm,clip]{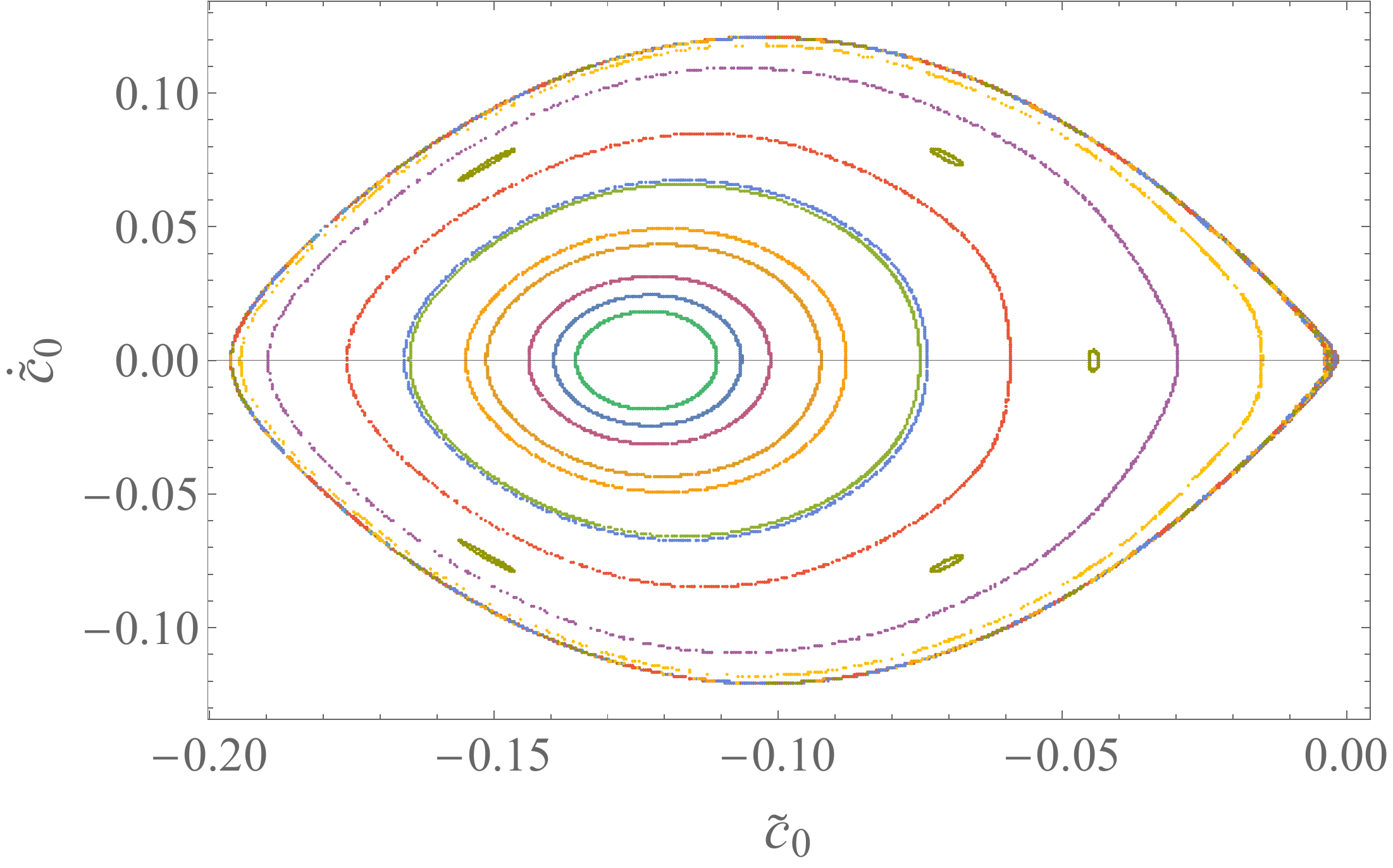}
\label{Fig:PP_E=9e-6}
}
\subfigure[Closeup of Fig.~\ref{Fig:PP_E=9e-6}]{
\includegraphics[width=8cm,clip]{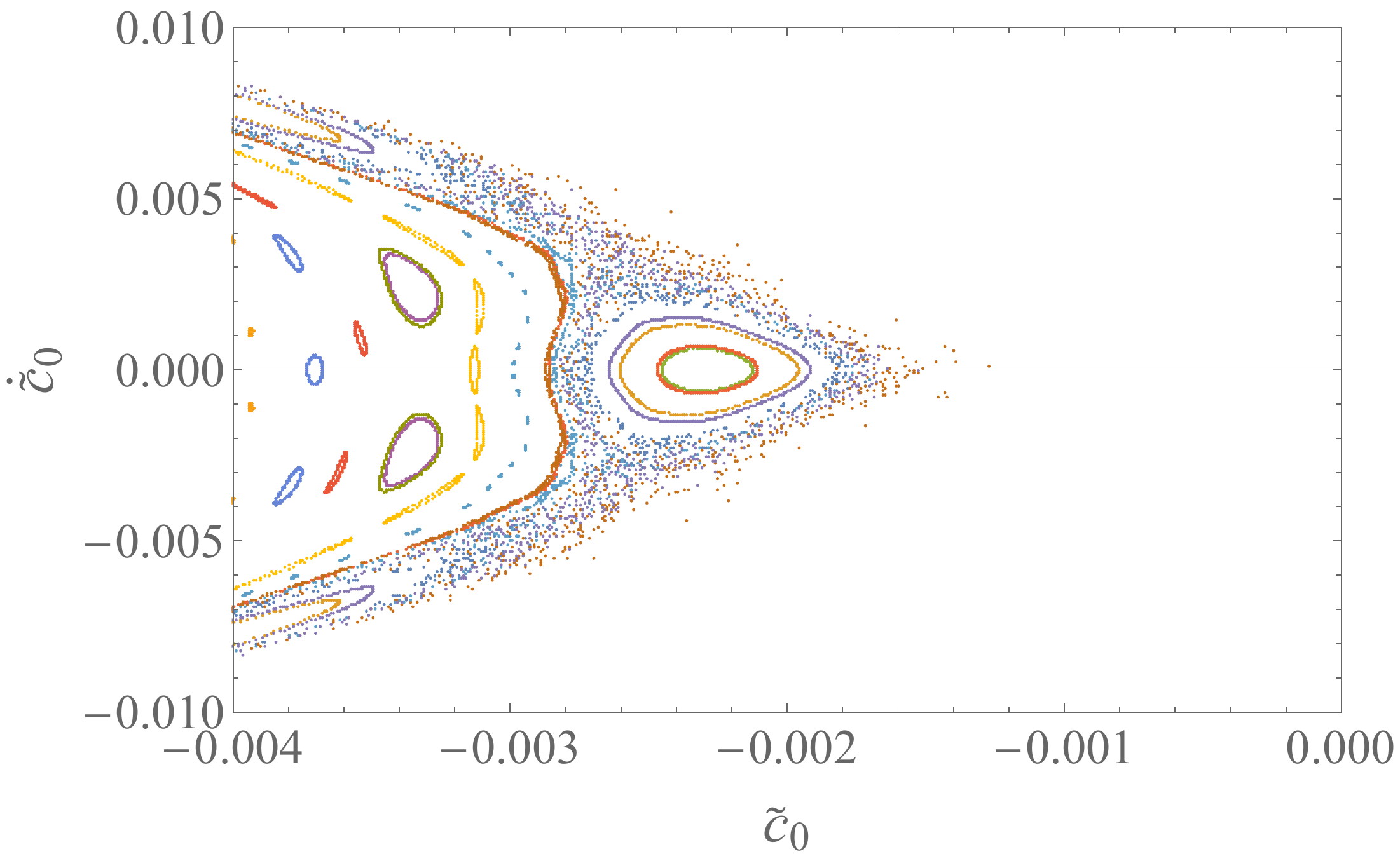}
\label{Fig:PP_E=9e-6_closeup}
}
\caption{
Poincar\'e plot obtained from the action (\ref{S_stabilized_deep}) for orbits with energy $E=9.28 \times 10^{-6}$ and $0<t<8000$. Though the most of the orbits form regular tori, some orbits near the saddle point of the potential form scattered plot indicating chaos.}
\label{Fig:PP}
\end{figure}

Solving the time evolution based on the action~(\ref{S_stabilized_deep}), we can examine if the system exhibits chaos or not by constructing a Poincar\'e section.
In Figure~\ref{Fig:PP}, we show the Poincar\'e sections defined by $\tilde c_1(t)=0$ and $\dot{\tilde c}_1(t)>0$ for bound orbits within the trapping potential.
From the action (\ref{S_stabilized_deep}) we can check that the energy of a bound orbit is negative, since the value of the potential at the unstable saddle point is equal to zero.
In Figure~\ref{Fig:PP}, we show the Poincar\'e plot for the orbits with energy $E=9.28 \times 10^{-6}$ and $0<t<8000$, where points with different colors corresponds to the numerical data of orbits for different initial data.
Most of the orbits form regular tori, while we can also see that some tori near the saddle point of the potential collapses to become scattered plots.
This saddle point corresponds to the black hole horizon, hence we confirmed that a black hole horizon works as a source of chaos.
Evaluating the Lyapunov exponent for these chaotic orbits, we find $\lambda \simeq 0.04 \times 2\pi T_H$, which satisfies the bound~(\ref{MSS}).
In section~\ref{sec:nonlinear} we study fully nonlinear motion of the string, and find that its Lyapunov exponent is of the same order as this value.

%

%
%


\section{Chaos of nonlinear dynamics of string}
\label{sec:nonlinear}

In this section, we demonstrate chaos of the string by full numerical simulations
to examine if the Lyapunov exponent of the chaotic motion obeys the bound (\ref{MSS}).
For this purpose we employ the numerical method to study dynamical string in AdS developed in Refs.~\cite{Ishii:2015wua,Ishii:2015qmj}.
To induce the nonlinear dynamics of the string, we ``quench'' the position of the string endpoints on the AdS boundary. Holographically it would correspond to the nonlinear dynamics of the gluon flux tube induced by the motion of the quark-anti quark pair.


To facilitate the numerical calculation, 
we introduce the ingoing Eddington-Finkelstein coordinates,
in which the metric (\ref{metric}) is given by 
\begin{equation}
 ds^2=\frac{R^2}{z^2}[-f(z)dV^2-2dVdz+d\vec{x}^2]
+ R^2 d\Omega_5^2
\ ,
\label{AdSBH2}
\end{equation}
where $z=R^2/r$, $z_H=R^2/r_H$, $f(z)=1-z^4/z_H^4$ and 
$V=t+z_\ast(z)$. 
Here, we introduced tortoise coordinate $z_\ast(z)=-\int^z_0 dz'/f(z')$. 
As the worldsheet coordinates, we take double null coordinates $(u,v)$ and specify the string position as
$V=V(u,v)$, $z=Z(u,v)$, $\vec{x}=\vec{X}(u,v)$.
The condition on the induced metric $h_{ab}$ for $u,v$ to be null coordinates is given by $h_{uu}=h_{vv}=0$.
In the double null coordinates, the Lagrangian for the string motion is proportional to $h_{uv}$ and
the Nambu-Goto action~(\ref{acF}) becomes
\begin{multline}
 S=-
  R^2\mathcal{T}\int dudv\, \frac{1}{Z^{2}}[-f(Z)V_{,u}V_{,v}
\\
-V_{,u}Z_{,v}-V_{,v}Z_{,u}+\vec{X}_{,u}\cdot \vec{X}_{,v}]\ .
\label{action1}
\end{multline}
From the action, we obtain the evolution equations of the string as
\begin{equation}
\begin{split}
&V_{,uv}=-\left(\frac{f}{Z}-\frac{f'}{2}\right)V_{,u}V_{,v}+\frac{1}{Z}\vec{X}_{,u}\cdot \vec{X}_{,v}\ ,\\
&Z_{,uv}=\left(\frac{f}{Z}-\frac{f'}{2}\right)(fV_{,u}V_{,v}+V_{,u}Z_{,v}+V_{,v}Z_{,u})
\\
& \qquad\quad
+\frac{2}{Z}Z_{,u}Z_{,v}-\frac{f}{Z}\vec{X}_{,u}\cdot \vec{X}_{,v}\ ,\\
&\vec{X}_{,uv}=\frac{1}{Z}(\vec{X}_{,u}Z_{,v}+\vec{X}_{,v}Z_{,u})\ ,
\end{split}
\label{evol}
\end{equation}
where $f'=df/dz$.

Hereafter, we take the unit $z_H=1$, in which the Hawking temperature is $T_H=1/\pi$. 
(Because the AdS scale $R$ appears just as a factor of the metric~(\ref{AdSBH2}), its value does not effect the string dynamics.)
Using the residual coordinate freedom, $u\to F(u)$ and $v\to G(v)$, we put boundaries of the worldsheet at $u-v=\pm\pi/2$.
As the initial condition, we use a static string shown in Fig.~\ref{Fig:static}.
For fixed $z_H=1$, 
the series of static solutions can be regarded as a one parameter family of $z_\textrm{center}$, 
where $z_\textrm{center}$ represents the $z$-coordinate  at the tip of the static string.
We pick up the static solution with $z_\textrm{center}=0.5$ as initial data.
Then, the separation of string endpoints is given by  $L=0.591$.

To induce the nonlinear motion of the string, 
we impose time dependent boundary conditions for string endpoints.
Introducing time and spatial coordinates on the worldsheet as $\tau=u+v$ and $\sigma=u-v$, we consider the ``quench'' of positions of the string endpoints along the transverse direction:  
\begin{equation}
 X_2(\tau,\sigma=\pm \pi/2)=\epsilon\, \alpha(\tau;\Delta \tau)\ ,
\end{equation}
where
\begin{equation}
 \alpha(\tau;\Delta \tau)
  =
\begin{cases}
 \exp\left[2\left(4-\frac{\Delta\tau}{\tau}-\frac{\Delta\tau}{\Delta\tau-\tau}\right)\right] & (0<\tau<\Delta \tau)\\
0 & (\textrm{else})
\end{cases}
.
\end{equation}
One can check that this function is $C^\infty$ and has a compact support in $0\leq\tau\leq\Delta \tau$.
There are two parameters $\epsilon$ and $\Delta \tau$,  which represent amplitude and time scale of the quench.
Boundary conditions for the other variables are $X_1(\tau,\sigma=\pm \pi/2)=\pm L/2$, $X_3(\tau,\sigma=\pm \pi/2)=0$ and $Z(\tau,\sigma=\pm \pi/2)=0$, and
$V(\tau,\sigma=\pm\pi/2)$ are determined by constraints $h_{uu}=h_{vv}=0$.
For these boundary conditions, the string motion is symmetric under $x\to -x$.
%
\begin{figure}[t]
  \centering
  \subfigure[Early time evolution]
 {\includegraphics[scale=0.33]{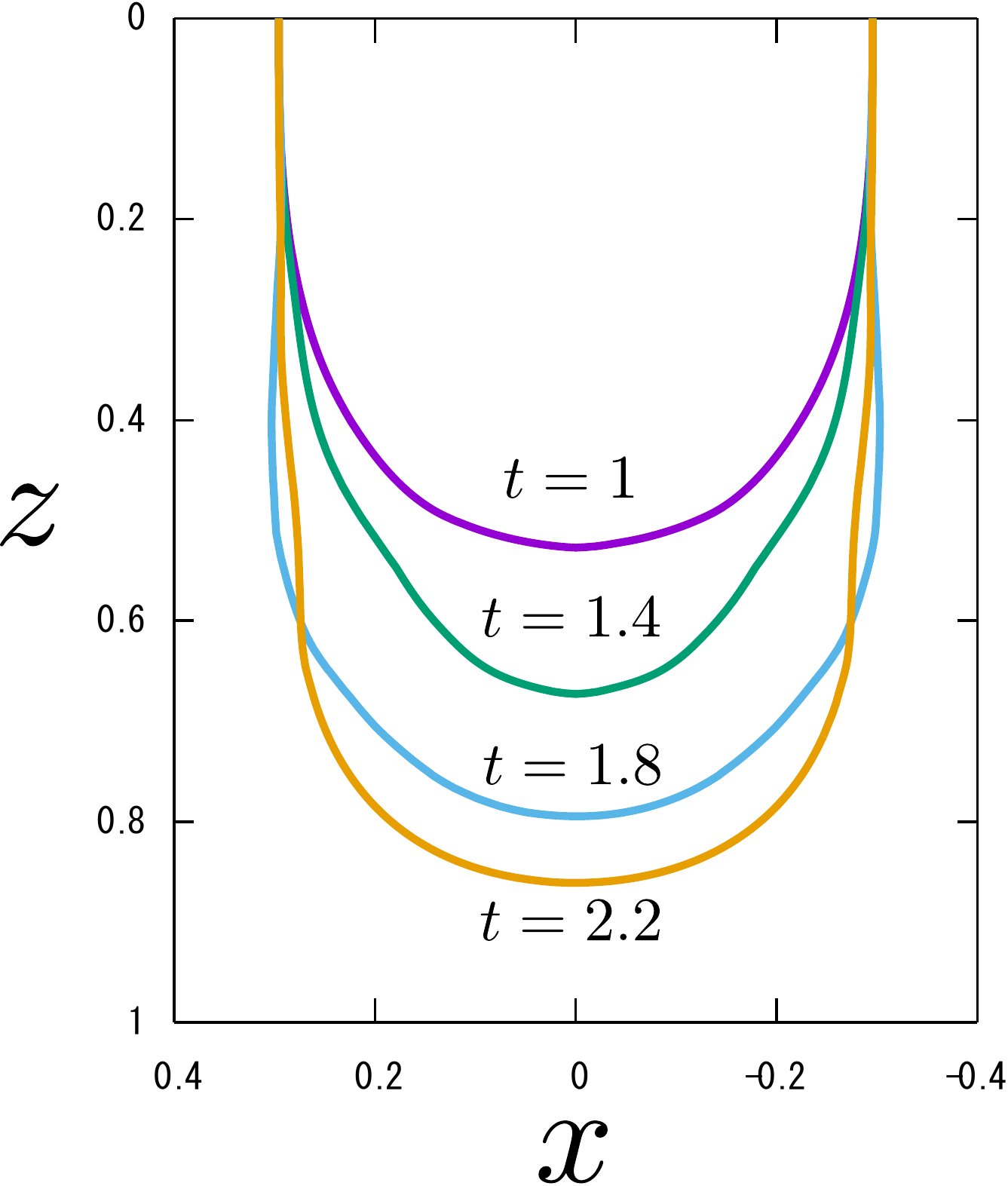}\label{early}
  }
  \subfigure[Late time evolution]
 {\includegraphics[scale=0.33]{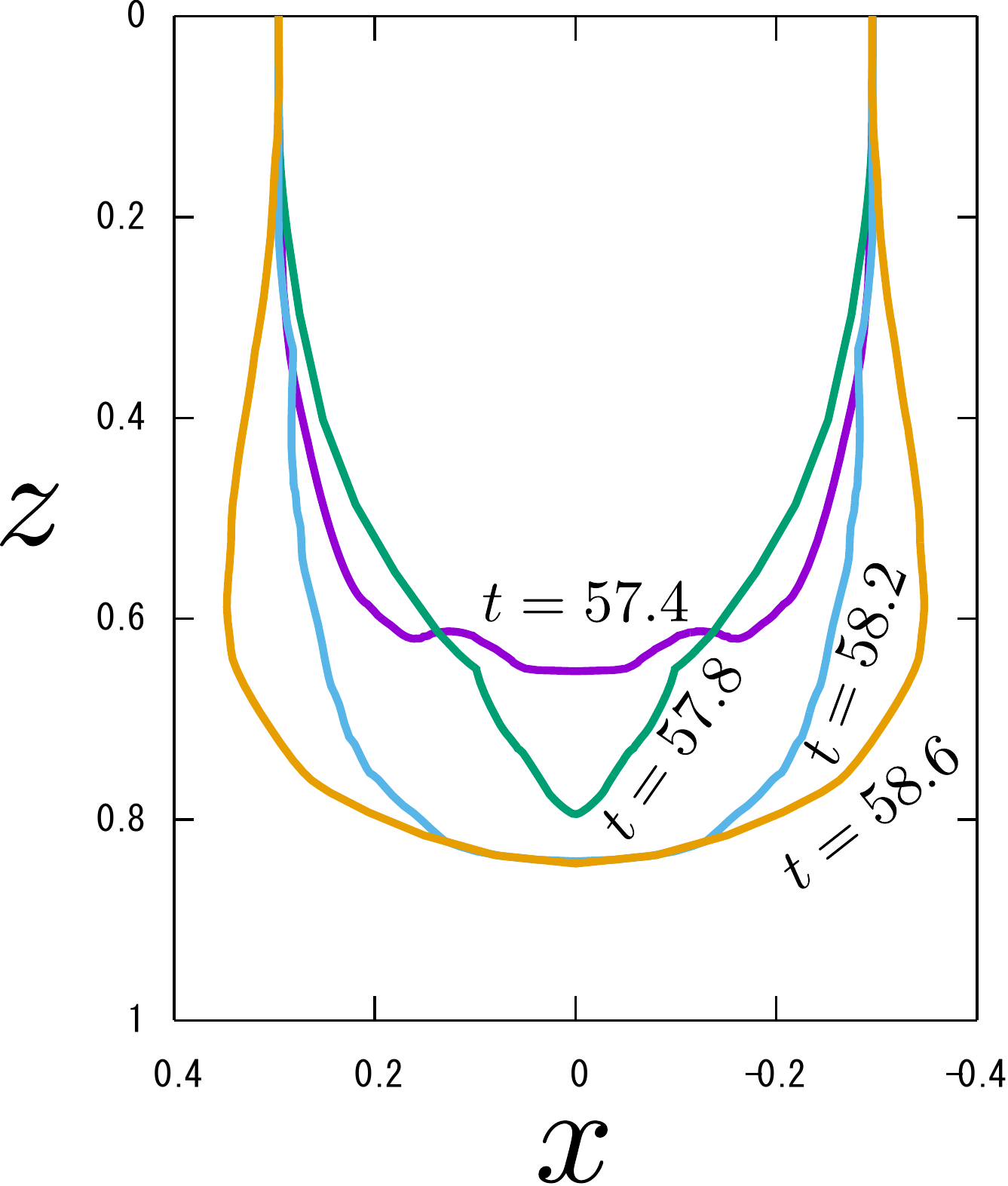}\label{late}
 }
 \subfigure[$z$-coordinate at center]
 {\includegraphics[scale=0.35]{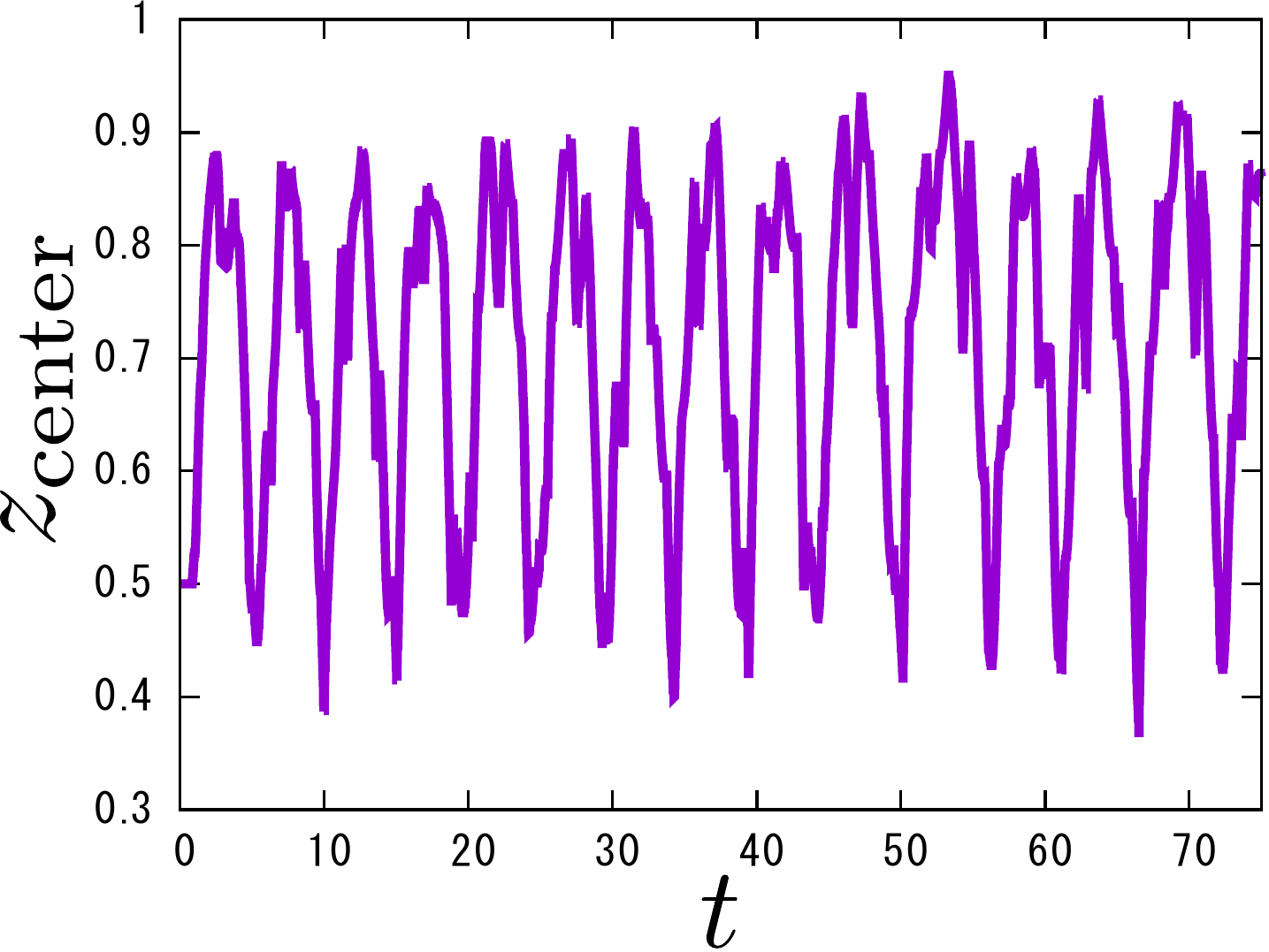}\label{zc}
  }
 \caption{
 Dynamical string for
 $\epsilon=0.030013$ and $\Delta \tau=2$.
 }
 \label{chaosFull}
\end{figure}

Fig.~\ref{chaosFull} shows the time evolution of the string in $(x,z)$-plane for $\epsilon=0.030013$ and $\Delta \tau=2$.
In this figure, we took the time slice of bulk coordinates: $t=V(u,v)+z_\ast(Z(u,v)))$.
The left and center panels are snapshots of the string for early ($1\leq t \leq 2.2$) and late time ($57.4\leq t \leq 58.6$).
(Although the string is also moving along $x_2$-direction, we do not show it for visibility.)
At the early time, the string profile seems smooth. On the other hand, at the late time, we can observe
small spatial fluctuations.
This would be reflecting the energy cascade on the string worldsheet in non-integrable geometries~\cite{Ishii:2016rlk}.
The right panel shows time dependence of the $z$-coordinate at the tip of the string.
To check the numerical accuracy, we monitored violation of constraints $h_{uu}=h_{vv}=0$.
The violation was  $\sim 10^{-3}$ compared to the typical scale of the dynamical solution.

To examine the sensitivity of the string motion to the initial conditions, we consider the linear perturbation
$V\to V+\delta V$, $Z\to Z+\delta Z$ and $\vec{X}\to \vec{X}+\delta \vec{X}$.
We numerically solve the linear equations for ($\delta V$, $\delta Z$, $\delta \vec{X}$) on the time dependent background.
Initial conditions are zero for all variables.
Boundary conditions are
$\delta X_2(\tau,\sigma=\pm \pi/2)= \alpha(\tau;\Delta\tau=2)$ and
$(\delta X_1,\delta X_3,\delta Z)|_{\sigma=\pm \pi/2}=0$. 
Again, $\delta V(\tau,\sigma=\pm\pi/2)$ are determined by linearized constraints $\delta h_{uu}= \delta h_{vv}=0$.

\begin{figure}[tb!]
  \centering
  \subfigure[$\epsilon=0.030013$]
 {\includegraphics[scale=0.45]{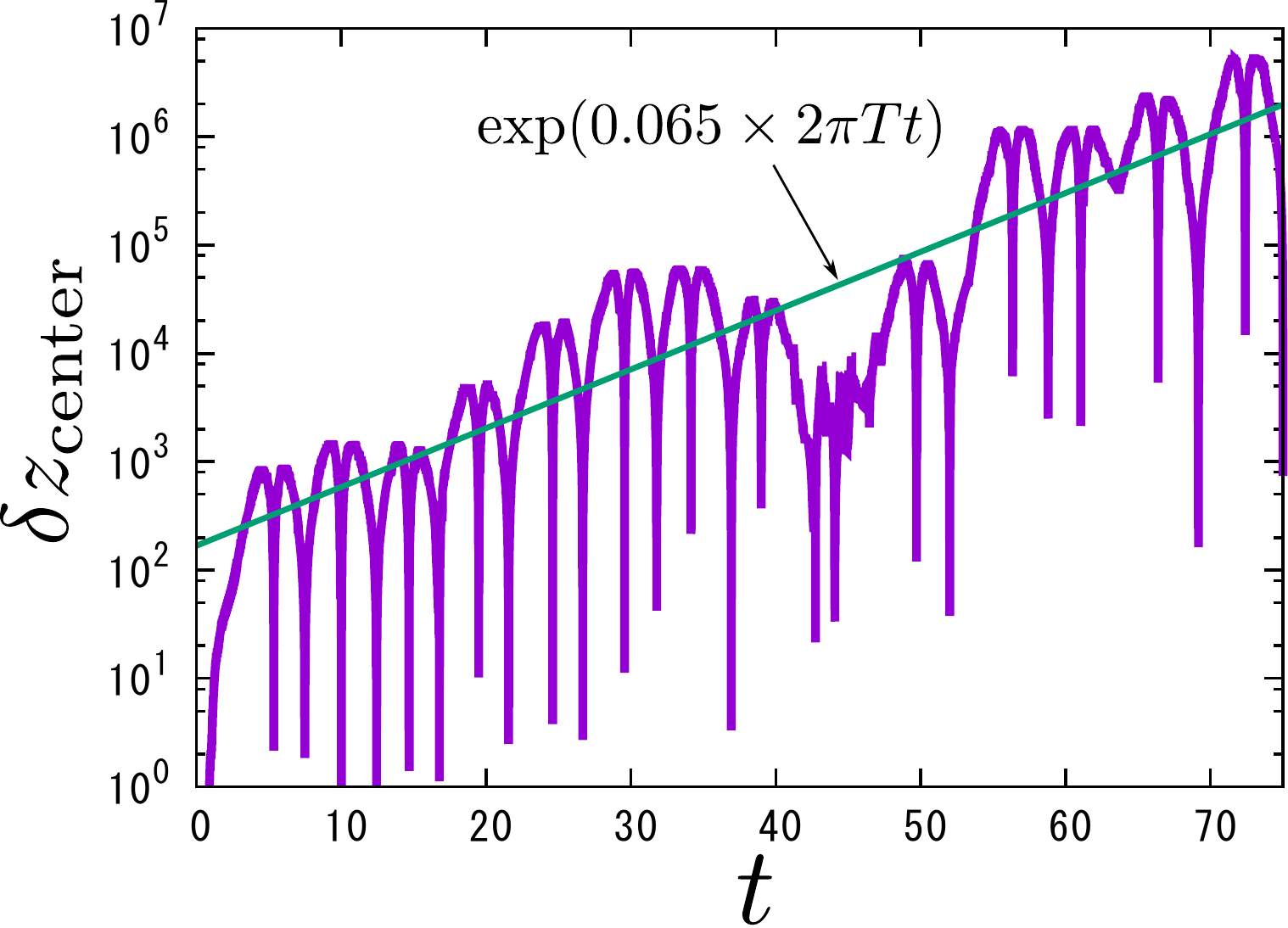}\label{dzc_chaos}
  }
  \subfigure[$\epsilon=0.029$]
 {\includegraphics[scale=0.45]{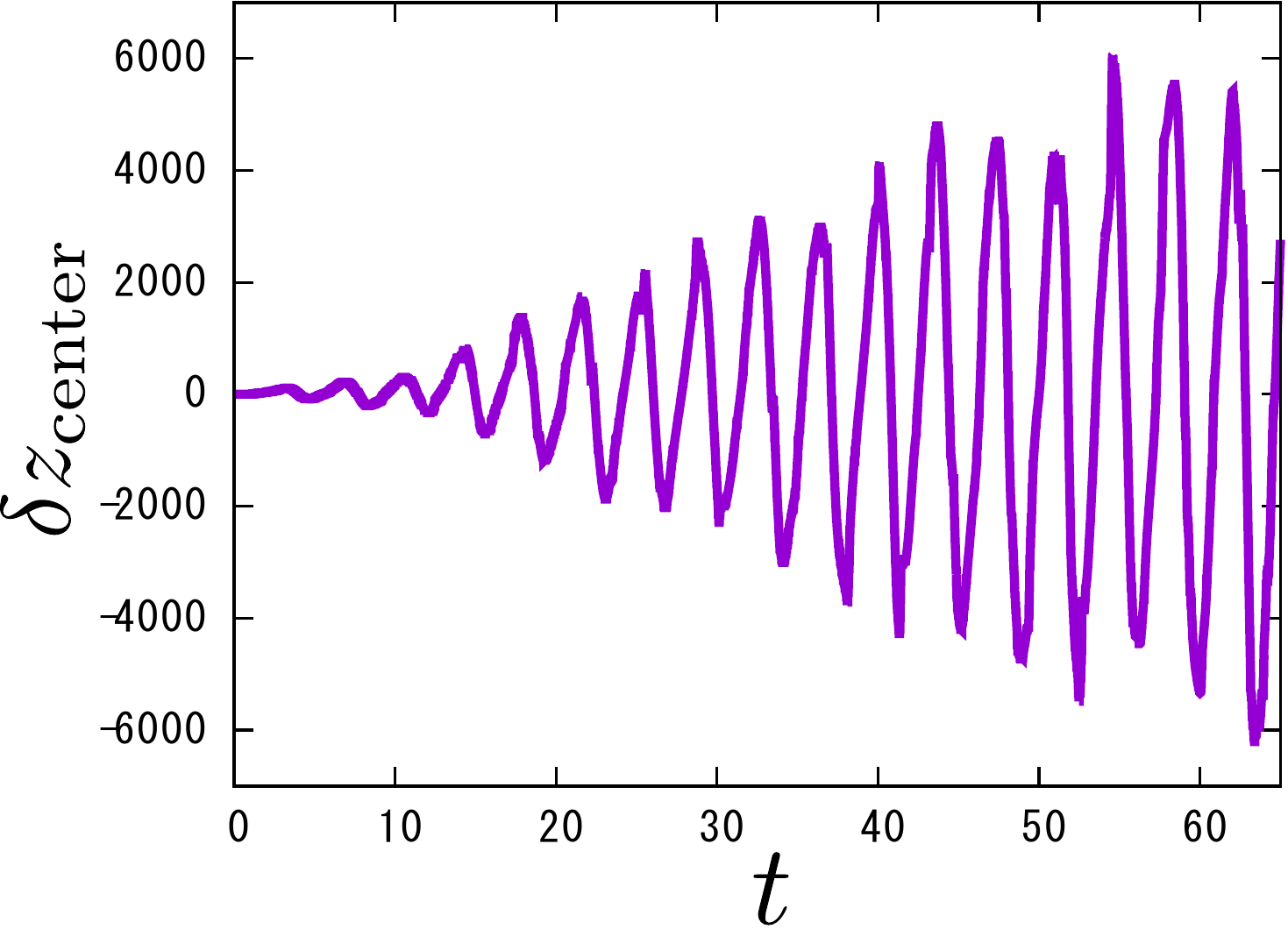}\label{dzc_lin}
 }
 \caption{
 Sensitivity of the string tip position $\delta z_\text{center}$ to initial conditions.
 }
 \label{dzcs}
\end{figure}

The results of the time evolution of the linear perturbation is summarized as follows.
Fig.~\ref{dzc_chaos} shows the time evolution of $\delta Z$ at the center of the string for $\epsilon=0.030013$.
As the horizontal axis, we took the bulk time coordinate $t=V(\tau,\sigma=\pm\pi/2)+z_\ast(Z(\tau,\sigma=\pm\pi/2))$.
We can see an exponential growth of the oscillation amplitude that demonstrates chaos.
Fitting the amplitude, we obtain the positive Lyapunov exponent as $\lambda\simeq 0.125 = 0.0625\times 2\pi T_H$.
This is smaller than the Lyapunov bound $2\pi T_H$.

\begin{figure}[htb!]
 \centering
\includegraphics[width=7cm]{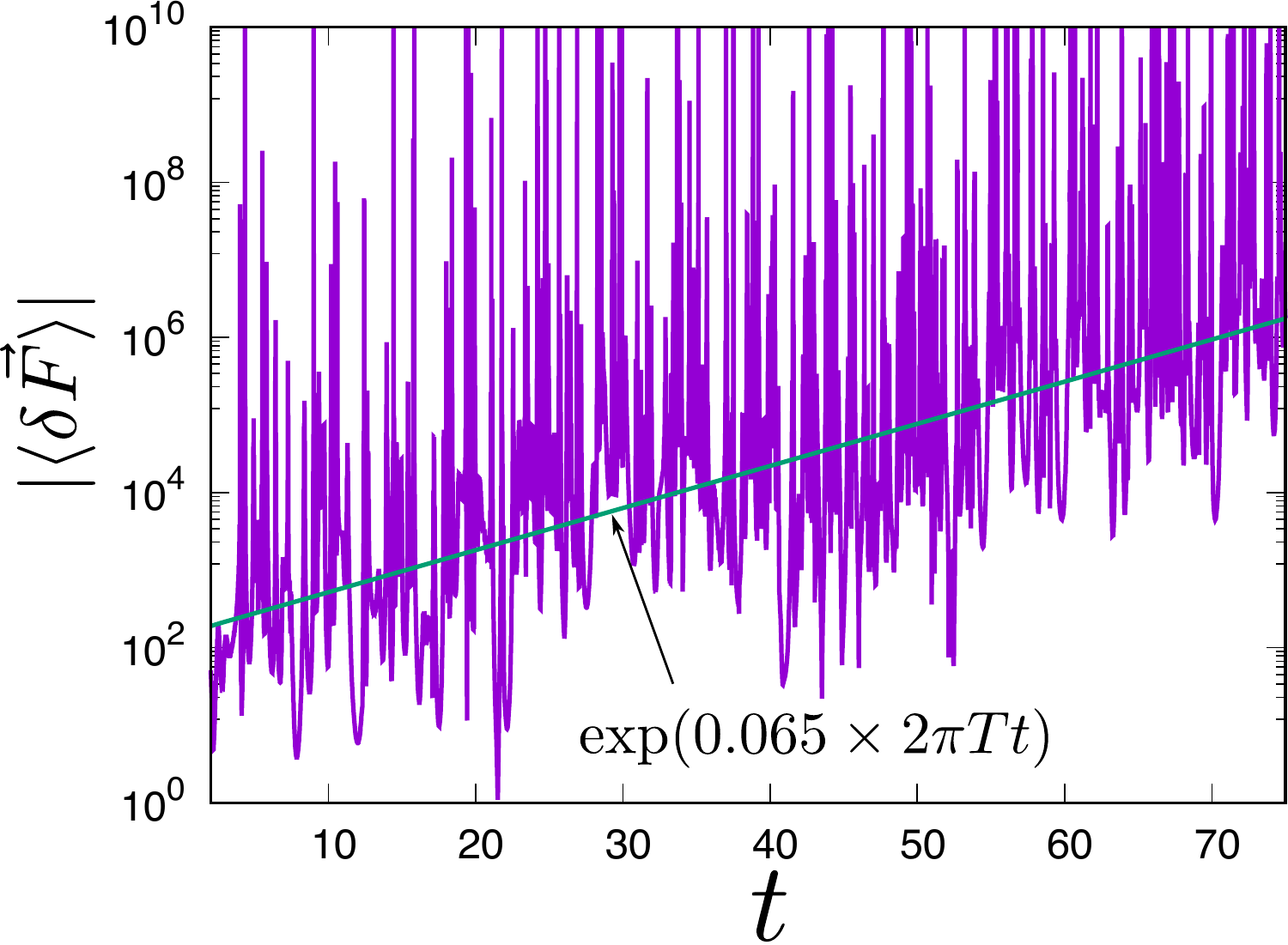}
\caption{Sensitivity of the force acting on the quark in the boundary field theory.
}
\label{Fig:delta_Force}
\end{figure}

The Lyapunov exponent can be measured also from an observable of the boundary field theory.
As shown in Ref.~\cite{Ishii:2015wua}, when the string endpoint is not moving, 
the force acting on the quark is given by
\begin{equation}
 \langle \vec{F}(t)\rangle  = \frac{\sqrt{\lambda_\textrm{'t Hooft}}}{4\pi}\partial_z^3 \vec{X}|_{z=0}\ ,
\end{equation}
where $\lambda_\textrm{'t Hooft}=R^4/\alpha'{}^2$ is the 't Hooft coupling.
Figure~\ref{Fig:delta_Force} shows the perturbation of the force $\langle \delta \vec{F}\rangle $.
Again, we took the bulk time coordinate for the horizontal axis.
We have dropped $\sqrt{\lambda_\textrm{'t Hooft}}$ in this plot.
$\langle\delta \vec{F}\rangle$ grows exponentially and its Lyapunov exponent is consistent with that from $\delta z_\text{center}(t)$.
We find chaos of the interquark forces, and it is a holographic prediction for the force in thermal gauge theories
with the AdS-Schwarzschild gravity dual. 

Some comments about the numerical simulations are in order.
We see that $\langle\delta \vec{F}\rangle$ is more spiky than $\delta z_\text{center}$ as functions of $t$.
This spiky profile is caused by cusps created on the string. 
In Ref.~\cite{Ishii:2015wua}, it is shown that the force on the quark diverges when the cusp arrives at the AdS boundary.

We can see the chaotic behaviour only in a small region of the parameter space of $\epsilon$
as one can see in Fig.~\ref{dzc_lin}.
In this figure, 
the amplitude of the quench is slightly decreased to $\epsilon=0.029$.
Then we see that the amplitude ceases to grow exponentially (Note that the scale of the vertical axis in Fig.~\ref{dzc_lin} is linear while that in Fig.~\ref{dzc_chaos} is log).
On the other hand, if we increase $\epsilon$ to $\epsilon\gtrsim 0.03002$, the string plunges into the black hole at finite time.

\section{Summary and discussions}

In this paper, we performed a nonlinear numerical simulation of time evolution of a suspended Nambu-Goto string
in the AdS-Schwarzschild background. In Sec.~\ref{sec:nonlinear}
we discovered that the force measured at the asymptotic boundary 
of the string has an exponential sensitivity against small difference in the initial conditions. We conclude that
the quark-antiquark force is chaotic in thermal gauge theories.

A perturbative analysis given in Sec.~\ref{sec:pert} shows 
collapse of regular tori of phase space orbits and also a positive Lyapunov exponent,
as expected.
The origin of the chaos should be the black hole horizon, and in Sec.~\ref{sec:universal}
our toy string model finds an unstable saddle near the horizon, characterized universally by the surface gravity $2\pi T_{\rm H}$.

The important point in our work is that the chaotic string motion in the bulk has a definite CFT interpretation:
the Wilson loop, or equivalently, the interquark forces. 
The exponential growth is physically provided by
\eqref{deltaF12} with the Lyapunov exponent $\lambda$. The holography calculates $\left\langle F(t)\right\rangle$
at large $N_c$ and at strong coupling limit, and the measured Lyapunov exponent is shown to satisfy
the universal chaos bound \eqref{MSS}, as seen in Sec.~\ref{sec:nonlinear}. This is our holographic prediction
for thermal gauge theories.

The chaos near the black hole horizon has been often
discussed in terms of the out-of-time (OTO) ordered correlators, and here we provide a relation to them.
We expect that the exponential growth would be
calculated by the following OTO correlator in the ${\cal N}=4$ supersymmetric Yang-Mills theory (SYM),
\begin{align}
&
C_T(t)
\nonumber 
\\
&
\equiv \left\langle 
{\rm Ptr} \exp\left[i\! \oint \!A_\mu dx^\mu \right] 
W(t) F_{01}(0)W(t)F_{01}(0)
\right\rangle_{\! T}
\label{OTOC}
\end{align}
This is an OTO correlator in the SYM, where the ordering of the operators
$W(t)$ and $F_{01}(0)$ is not time-ordered. The first part is the standard Wilson loop operator
with a rectangular shape (the spatial separation is $L$ along $x^1$ direction 
while the time length is $T\to\infty$).
$W(t)$ is a gauge-invariant operator inserted at $t=0$, which could be, for example,
a spatial integration of the energy momentum tensor, $\int \! d^3x\,  T_{00}(t)$.
The field strength $F_{01}(t=0)$, or a Plaquette, 
is inserted on the path of the Wilson loop at $t=0$,
and basically it amounts to an infinitesimal deformation of the path of the Wilson loop 
at the location.

The reason why we expect $C_T(t)$ in \eqref{OTOC}
can detect the exponential growth of the unstable
string fluctuation is as follows.
In \cite{Shenker:2013pqa} and the subsequent study the OTO
correlation function of the form $\langle W(t)V(0)W(t)V(0)\rangle_T$ was shown to
possess an exponential growth in time, where $W(t)$ is an operator perturbing 
the system while $V(0)$ is another operator for which one wants to evaluate
the correlation function. In the gravity dual, the operator $W$ perturb the  
background geometry with a horizon. A time-ordered correlation function
should approach a factorized $\langle W(t)W(t)\rangle \langle V(0)V(0)\rangle$, 
while the OTO correlation function should decay with the Lyapunov
exponent. In our case the operator $V$ should correspond to an infinitesimal
fluctuation of the Wilson loop, as the Wilson loop itself is a background on which
the correlation function is measured. And the operator $W$ should be 
the one perturbing 
the system,
so it can be inserted anywhere.
In the gravity dual, we have found that the Wilson loop fluctuation can have
an exponential growth near the horizon, thus a natural OTO correlator which we 
can suggest is \eqref{OTOC}~\footnote{
Note that the OTOC defined in \cite{Maldacena:2015waa} uses a peculiar insertion 
of divided thermal ensemble operators $\exp[-\beta H/4]$, thus our chaotic $\langle \delta F(t)\rangle$
may not be written in the form of OTOC defined in \cite{Maldacena:2015waa}.}.

The unstable string configuration plays a role of instanton amplitudes in string breaking.
The unstable configuration is a saddle point dividing the connected and broken
string configurations, thus the Lyapunov exponent is expected to be relevant
to the instanton amplitude at the string breaking transition. The universality
seen in this paper is important in showing that the breaking amplitude in the WKB
approximation should be given by 
\begin{align}
\Gamma \sim \exp \left[- c LT\right]
\end{align}
with some positive numerical coefficient $c$, since we can use only the temperature
for fitting the dimension of the exponent. The coefficient $c$ can be calculated
using our Lyapunov exponent.

The string breaking is known to be associated with thermal entropy production 
\cite{Kharzeev:2014pha,Hashimoto:2014fha,Iatrakis:2015sua,Fadafan:2015ynz},
which is argued to be relevant for quarkonium dissociation. 
On the other hand, the Lyapunov exponent provides Kolmogorov-Sinai entropy
production rate, and we expect that these two issues are related with each other.

\acknowledgments
K.H.\ would like to thank T.~Hatsuda, Y.~Hidaka and M.~Hongo for valuable comments.
The work of K.H.\ was supported in part by JSPS KAKENHI Grants No.~JP15H03658, 
No.~JP15K13483, and No.~JP17H06462. 
The work of N.T.\ was supported in part by JSPS KAKENHI Grant Number 16H06932.

\appendix

\section{Universality in a $(p,q)$-string}
\label{App:pq-string}

In section~\ref{sec:universal}, we observed that the Lyapunov exponent for a fundamental string takes a universal value that depend only on the horizon temperature but independent on the other physical parameters.
In this appendix, let us consider a more generic situation in string
theory to examine the universality of the Lyapunov exponent. 

We consider a $(p,q)$-string in the near-horizon geometry of D3-F1-D1 bound state with
a temperature. For the action of the $(p,q)$-string action, we use the S-duality covariant action
\cite{Schmidhuber:1996fy}
\begin{align}
S = &
-\int dtd\sigma \left[
\sqrt{q^2 e^{-2\phi}+(p-q\chi)^2} \sqrt{-\det G[X]}  
\right.
\nonumber \\
& \hspace{20mm}
\left.
+(q A_{t\mu} +p B_{t\mu})\frac{dX^\mu}{d\sigma}
\right] \, .
\end{align}
Here $A$ and $B$ are Ramond-Ramond (RR) and NS-NS 2-form, and
$\phi$ and $\chi$ are the dilaton and the RR 0-form axion, respectively.
The near-horizon geometry of the D3-F1-D1 bound state is given in \cite{Cai:2000yk},
\begin{align}
ds^2 &= F^{-1/2}
\left(
h'(-f dt^2+ dx_1^2) + h(dx_2^2+dx_3^2)\right) 
\nonumber
\\
& \quad + F^{1/2} f^{-1} dr^2, 
\\
e^{2\phi} & = g_s^2 h h', 
\\
\chi &= \frac{-1}{g_s F}\tan\varphi \sin\theta, 
\\
B_{01} & =
H^{-1} \coth\alpha \sin\varphi, 
\\
A_{01} & = 
-H^{-1} \coth\alpha\sin\varphi, 
\end{align}
with the following definitions of the functions,
\begin{align}
f &\equiv 1-\frac{r_H^4}{r^4}, \quad H \equiv \frac{r_H^4\sinh^4\alpha}{r^4}\, , \quad
h\equiv F/G \, ,
\nonumber
\\
h' &\equiv F/H\, , 
F\equiv 1 + \cos^2\varphi \frac{r_H^4}{\sinh^2\alpha}{r^4}\, , 
\nonumber \\
G& \equiv 1 + \cos^2\varphi \cos^2\theta \frac{r_H^4 \sinh^2\alpha}{r^4}\, .
\end{align}
The background geometry is quite involved, but after substituting
the background into the $(p,q)$-string action, we obtain,  for 
the rectangular shape,
\begin{align}
S =& -\int \! dt
\left[
L c_1\sqrt{r_0-r_H} \sqrt{1-\frac{c_2}{(r_0-r_H)^2}\dot{r}_0^2}
\right.
\nonumber \\
&
\quad\quad
\left.-(c_3 + c_4 L)(r_0-r_H)\right] \, .
\end{align}
Here, $c_i ~(i=1,2,3,4)$ is a constant ($c_1$, $c_2$ and $c_3$ are positive).
It is a function of the parameters of the background geometry, $(\alpha, \varphi, \theta, g_2)$, and also of $(p,q)$.
With this action, we find the maximum of the potential energy. As before,
we expand the
action around that point and also expand it to the leading order in the kinetic term, then we
obtain
\begin{align}
S = \frac{-c_1^2 L^2}{4(c_3\!+\!c_4 L)}
+ \frac{4c_2 (c_3\!+\!c_4 L)^3}{c_1^2 L^2}
\left[
\left(\frac{d}{dt}\delta r\right)^2 + \frac{1}{4c_2} (\delta r)^2
\right] \, .
\end{align}
Let us evaluate the coefficient $c_2$ explicitly. It is given by
\begin{align}
\frac{c_2}{(r_0-r_H)^2} = \frac{F}{h' f^2} \,  
\end{align}
and at the near-horizon region $r_0 \sim r_H$, we find
\begin{align}
c_2 = \frac{r_H^2 \cosh^2\alpha}{16}\, .
\end{align}
The corrections, which are of order ${\cal O}(L)$, can be ignored
for our interest of small $L$.
Noting that the Hawking temperature of this generic geometry is given by
\begin{align}
T_{\rm H} = \frac{1}{\pi r_H \cosh\alpha}, 
\end{align}
we find that the action at the top of the potential hill for slow motion of the
rectangular $(p,q)$-string is 
\begin{align}
S = \frac{4c_2 (c_3\!+\!c_4 L)^3}{c_1^2 L^2}
\left[
\left(\frac{d}{dt}\delta r\right)^2 + (2\pi T_{\rm H})^2 (\delta r)^2
\right] \, .
\label{effst2}
\end{align}
This has the same form as the previous case \eqref{effst}.
The dynamics is completely governed by the Hawking temperature, again.
So the Lyapunov exponent $\lambda$ for the motion of the $(p,q)$-string
in the near horizon geometry of the D3-D1-F1 bound state background is
given again by
\begin{align}
\lambda = 2\pi T_{\rm H} \, .
\end{align}
Hence the value of the Lyapunov exponent is universal, and is given by the Hawking
temperature
even in this generalized setup.

\section{Derivation of effective action}
\label{App:Seff}

In this appendix, we show the details of the derivation of the perturbative action~(\ref{S_stabilized_deep}) studied in section~\ref{sec:pert}.
As explained there, we parameterize the perturbation of a static string using the proper distance along the normal vector of the string trajectory defined by Eqs.~(\ref{normal}), (\ref{deformation}) and Fig.~\ref{Fig:n-gauge}.

The action for the perturbative motion
is derived by
expanding the metric functions in (\ref{metric2}) as
\begin{align}
g_\mu(r) &= 
  g_\mu ( r_\text{BG} )
+ g'_\mu( r_\text{BG} )  \xi n^r 
\notag \\
&
+ \frac{ g''_\mu( r_\text{BG} ) }{2} (\xi n^r )^2
+ \frac{ g'''_\mu( r_\text{BG} )}{6}  (\xi n^r )^3
+ \cdots,
\end{align}
and using the line elements given by
\begin{align}
dx &=
\left(  x_\text{BG}' +  (n^x)' \xi + n^x \xi' \right) d\ell
+ n^x \dot\xi dt
,
\\
dr &=
\left( r_\text{BG}' +  (n^r)' \xi + n^r  \xi' \right) d\ell
+ n^r  \dot\xi dt,
\end{align}
where a dot denotes a $t$ derivative.

Up to a surface term, the quadratic action is given by
\begin{equation}
\frac{S^{(2)}}{\cal T}
=
\int dt
\int_{-\infty}^\infty d\ell
\left[
C_{tt}(\ell)
\dot\xi^2
- C_{\ell\ell}(\ell)\xi'^2
- C_{00}(\ell) \xi^2
\right],
\label{L2}
\end{equation}
where
\begin{equation}
\begin{aligned}
C_{tt}(\ell) &= \frac{r_\text{BG}(\ell)}{2\sqrt{r_\text{BG}^4(\ell)-1}},
\quad
C_{\ell\ell}(\ell) = 
\frac{\sqrt{r_\text{BG}^4(\ell)-1}}{2r_\text{BG}(\ell)},
\\
C_{00}(\ell)&=
\frac{r_0^4 + 3 r_\text{BG}^4 (\ell)- 4r_0^4 r_\text{BG}^4(\ell) - r_0^4 r_\text{BG}^8(\ell) + r_\text{BG}^{12}(\ell)}{r_\text{BG}^5(\ell) \left(r_\text{BG}^4(\ell)-1\right)^{3/2}}.
\end{aligned}
\label{coeffs}
\end{equation}
Then the perturbation equation is given by
\begin{equation}
\omega^2 C_{tt}(\ell) \xi(\ell) + 
\partial_\ell\left[ C_{\ell\ell}(\ell) \xi'(\ell)\right]
- C_{00}(\ell) \xi(\ell) = 0,
\label{xieq(App)}
\end{equation}
where we separated the variable as $\xi(t,\ell) = \xi(\ell) e^{i\omega t}$.
This equation is of the Sturm-Liouville form with the weight function
$W(\ell) = C_{tt}(\ell)$,
with which the inner product is defined as
\begin{equation}
\left(\xi,\zeta\right) \equiv \int_{-\infty}^\infty  W(\ell) \xi(\ell) \zeta(\ell)d\ell.
\end{equation}
Then we truncate the perturbation up to the first two lowest excitations as
\begin{equation}
\xi = c_0(t) e_0(\ell) + c_1(t) e_1(\ell),
\quad
\left( e_m,e_n \right) = \delta_{mn},
\label{xi-ansatz}
\end{equation}
where $e_{0,1}(\ell)$ are the mode functions with the lowest and the next lowest eigenfrequencies.
Note that $e_0(\ell)$ and $e_1(\ell)$ are even and odd functions of $\ell$, respectively.
Plugging (\ref{xi-ansatz}) into the quadratic Lagrangian (\ref{L2}), we find
\begin{equation}
\frac{S^{(2)}}{\cal T}
=
\int dt
\sum_{n=0,1}
\left(
 \dot c_n^2 - \omega_n^2 c_n^2
\right) .
\end{equation}
The cubic-order terms of the Lagrangian are given by
\begin{align}
\frac{{\cal L}^{(3)}}{\cal T} &=
C_0(\ell)\xi^3
+
C_1(\ell) \xi^2 \xi'
+ C_2(\ell) \xi  \xi'^2
+ C_3(\ell) \xi \dot \xi^2
\notag \\
&=
\left(
C_0(\ell) - \frac13 C_1'(\ell)
\right)
\xi^3
+ C_2(\ell) \xi  \xi'^2
+ C_3(\ell) \xi \dot \xi^2,
\end{align}
where the second equality 
holds up to a surface term.
Then the cubic order part of the action is given by
\begin{align}
&\frac{S^{(3)}}{\cal T} =
\int dt
\int_{-\infty}^\infty d\ell \Biggl\{
\left[\left( C_0 - \frac{C_1'}{3} \right) e_0^3 + C_2 e_0 e_0'\right] c_0^3(t)
\notag \\
&+ \left[
3 \left( C_0 - \frac{C_1'}{3} \right) e_0 e_1^2
+ C_2 \left(
e_0 e_1'^2 + 2 e_0' e_1 e_1'
\right)
\right]c_0(t)c_1^2(t)
\notag \\
&+ C_3
\left(
e_0^3 c_0 \dot c_0^2
+ e_0 e_1^2 c_0 \dot c_1^2 
+ 2 e_0 e_1^2 \dot c_0 c_1 \dot c_1
\right)
\Biggr\},
\end{align}
where we retained only even functions of $\ell$ in the integrand since the odd part does not give contribution after the integration.
We an evaluate the coefficients appearing in this action numerically.
For example, when the tip of the string is at $r=r_0 = 1.1$, the action is given by
\begin{multline}
\frac{S}{\cal T} = 
\int dt
\sum_{n=0,1}
\left(
 \dot c_n^2 - \omega_n^2 c_n^2
\right)
+ 11.3 c_0^3 
+ 21.5 c_0 c_1^2 
\\
+ 10.7 c_0 \dot c_0^2 
+ 3.32 c_0 \dot c_1^2 
+ 6.64 \dot c_0 c_1 \dot c_1,
\label{S_proper}
\end{multline}
where $\omega_0^2 = - 1.40$ and $\omega_1^2 = 7.57$.

The action (\ref{S_proper}) describes motion of the string in a trapping potential with an unstable equilibrium point on its edge, as explained above. One problem of this action is that the kinetic term of the Hamiltonian constructed from this action is not positive definite in some regions within the trapping potential, hence the time evolution there is not well-posed.
One way to ameliorate this problem is to 
apply the following variable change 
\begin{equation}
c_0 = \tilde c_0 + \alpha_1 \tilde c_0^2 + \alpha_2 \tilde c_1^2,
\quad
c_1 = \tilde c_1 + \alpha_3 \tilde c_0 \tilde c_1
\label{tildecdef}
\end{equation}
to the action and neglect ${\cal O}(\tilde c_i{}^4)$ terms, where $\alpha_i$ are constants.
By taking appropriate values for $\alpha_i$, we can ensure that the kinetic term is positive definite within the trapping potential.
An example of such a choice is $\alpha_1 = -1.5$, $\alpha_2 = -0.5$ and $\alpha_3=-1$, that gives
\begin{multline}
\frac{S}{\cal T} = \int dt
\sum_{n=0,1}
\left(
 \dot{\tilde c}_n^2 - \omega_n^2 {\tilde c}_n^2
\right)
+ 7.11 \tilde c_0^3
+ 35.3 \tilde c_0 \tilde c_1^2 
\\
+ 4.66 \tilde c_0 \dot {\tilde c}_0^2 
+ 1.32 \tilde c_0 \dot {\tilde c}_1^2 
- 7.57 \dot {\tilde c}_0 \tilde c_1 \dot {\tilde c}_1,
\label{S_stabilized_deep(App)}
\end{multline}
for which the time evolution within the trapping potential is well-posed.
This is the cubic order action (\ref{S_stabilized_deep}) studied in section~\ref{sec:pert}, and it exhibits chaos for orbits near the unstable maximum corresponding to the black hole horizon.

\vfill



\begin{thebibliography}{99}


\bibitem{Maldacena:1997re} 
  J.~M.~Maldacena,
  Int.\ J.\ Theor.\ Phys.\  {\bf 38}, 1113 (1999)
  [Adv.\ Theor.\ Math.\ Phys.\  {\bf 2}, 231 (1998)]
  [hep-th/9711200].







  
\bibitem{Shenker:2013yza} 
  S.~H.~Shenker and D.~Stanford,
  JHEP {\bf 1412}, 046 (2014)
  [arXiv:1312.3296 [hep-th]].
  

\bibitem{Shenker:2013pqa} 
  S.~H.~Shenker and D.~Stanford,
  JHEP {\bf 1403}, 067 (2014)
  [arXiv:1306.0622 [hep-th]].

  
\bibitem{Leichenauer:2014nxa} 
  S.~Leichenauer,
  Phys.\ Rev.\ D {\bf 90}, no. 4, 046009 (2014)
  [arXiv:1405.7365 [hep-th]].

\bibitem{Kitaev-talk}
A.~Kitaev, 
talk given at Fundamental Physics Symposium, Nov.~2014.


\bibitem{Shenker:2014cwa} 
  S.~H.~Shenker and D.~Stanford,
  JHEP {\bf 1505}, 132 (2015)
  [arXiv:1412.6087 [hep-th]].
  
\bibitem{Jackson:2014nla} 
  S.~Jackson, L.~McGough and H.~Verlinde,
  Nucl.\ Phys.\ B {\bf 901}, 382 (2015)
  [arXiv:1412.5205 [hep-th]].
  

\bibitem{Polchinski:2015cea} 
  J.~Polchinski,
  arXiv:1505.08108 [hep-th].


\bibitem{Maldacena:2015waa} 
  J.~Maldacena, S.~H.~Shenker and D.~Stanford,
  JHEP {\bf 1608}, 106 (2016)
  [arXiv:1503.01409 [hep-th]].


    
\bibitem{Larkin}
A.~I.~Larkin and Y.~N.~Ovchinnikov,
JETP 28, 6 (1969): 1200-1205.

\bibitem{Basu:2011dg}
P.~Basu, D.~Das and A.~Ghosh,
  ``Integrability Lost,''
  Phys.\ Lett.\ B {\bf 699} (2011) 388
  [arXiv:1103.4101 [hep-th]].
  

\bibitem{T11}
P.~Basu and L.~A.~Pando Zayas,
  ``Chaos Rules out Integrability of Strings in AdS$_5 \times T^{1,1}$,''  
Phys.\ Lett.\ B {\bf 700} (2011) 243 [arXiv:1103.4107 [hep-th]]; 
  ``Analytic Non-integrability in String Theory,''
  Phys.\ Rev.\ D {\bf 84} (2011) 046006
  [arXiv:1105.2540 [hep-th]].   

\bibitem{WQCD} 
 P.~Basu, D.~Das, A.~Ghosh and L.~A.~Pando Zayas,
  ``Chaos around Holographic Regge Trajectories,''
  JHEP {\bf 1205} (2012) 077
  [arXiv:1201.5634 [hep-th]]. \\ 
L.~A.~Pando Zayas and D.~Reichmann, 
``A String Theory Explanation for Quantum Chaos in the Hadronic Spectrum,'' 
JHEP {\bf 1304} (2013) 083 
[arXiv:1209.5902 [hep-th]]. 
 
\bibitem{D-brane}
A.~Stepanchuk and A.~A.~Tseytlin,
  ``On (non)integrability of classical strings in p-brane backgrounds,''
  J.\ Phys.\ A {\bf 46} (2013) 125401
  [arXiv:1211.3727 [hep-th]]. \\ 
   Y.~Chervonyi and O.~Lunin,
  ``(Non)-Integrability of Geodesics in D-brane Backgrounds,''
  JHEP {\bf 1402} (2014) 061
  [arXiv:1311.1521 [hep-th]].
  
\bibitem{complex-beta}  
D.~Giataganas, L.~A.~Pando Zayas and K.~Zoubos,
  ``On Marginal Deformations and Non-Integrability,''
  JHEP {\bf 1401} (2014) 129
  [arXiv:1311.3241 [hep-th]].  

\bibitem{NR}
  D.~Giataganas and K.~Sfetsos,
  ``Non-integrability in non-relativistic theories,''
  JHEP {\bf 1406} (2014) 018
  [arXiv:1403.2703 [hep-th]]. \\ 
 X.~Bai, J.~Chen, B.~H.~Lee and T.~Moon,
  ``Chaos in Lifshitz Spacetimes,'' 
   J.\ Korean Phys.\ Soc.\  {\bf 68} no.5 (2016)  639 
  [arXiv:1406.5816 [hep-th]].

\bibitem{T11-ppwave} 
  Y.~Asano, D.~Kawai, H.~Kyono and K.~Yoshida,
  ``Chaotic strings in a near Penrose limit of AdS$_{5} \times T^{1,1}$,''
  JHEP {\bf 1508} (2015) 060 
  [arXiv:1505.07583 [hep-th]].

\bibitem{gamma} 
  K.~L.~Panigrahi and M.~Samal,
  ``Chaos in classical string dynamics in $\hat{\gamma}$ deformed AdS$_5 \times T^{1,1}$,''
  arXiv:1605.05638 [hep-th].  
  
\bibitem{BH}
 L.~A.~Pando Zayas and C.~A.~Terrero-Escalante,
  ``Chaos in the Gauge / Gravity Correspondence,''
  JHEP {\bf 1009} (2010) 094
  [arXiv:1007.0277 [hep-th]].\\ 
P.~Basu, P.~Chaturvedi and P.~Samantray, 
``Chaotic dynamics of strings in charged black hole backgrounds,'' 
arXiv:1607.04466. 
  

\bibitem{AKY}
 Y.~Asano, H.~Kyono and K.~Yoshida,
  ``Melnikov's method in String Theory,''
 JHEP {\bf 1609} (2016) 103
  [arXiv:1607.07302 [hep-th]]. 

\bibitem{Ishii:2016rlk} 
  T.~Ishii, K.~Murata and K.~Yoshida,
  ``Fate of chaotic strings in a confining geometry,''
  Phys.\ Rev.\ D {\bf 95}, no. 6, 066019 (2017)
  doi:10.1103/PhysRevD.95.066019
  [arXiv:1610.05833 [hep-th]].


\bibitem{Hashimoto:2016dfz} 
  K.~Hashimoto and N.~Tanahashi,
  Phys.\ Rev.\ D {\bf 95}, 024007 (2017)
  [arXiv:1610.06070 [hep-th]].

\bibitem{Ishii:2015wua} 
  T.~Ishii and K.~Murata,
  ``Turbulent strings in AdS/CFT,''
  JHEP {\bf 1506}, 086 (2015)
  [arXiv:1504.02190 [hep-th]].

\bibitem{Ishii:2015qmj} 
  T.~Ishii and K.~Murata,
  ``Dynamical AdS strings across horizons,''
  JHEP {\bf 1603}, 035 (2016)
  [arXiv:1512.08574 [hep-th]].




%
%


\bibitem{Maldacena:1998im} 
  J.~M.~Maldacena,
  Phys.\ Rev.\ Lett.\  {\bf 80}, 4859 (1998)
  [hep-th/9803002].
  
\bibitem{Rey:1998ik} 
  S.~J.~Rey and J.~T.~Yee,
  Eur.\ Phys.\ J.\ C {\bf 22}, 379 (2001)
  [hep-th/9803001].

\bibitem{Rey:1998bq} 
  S.~J.~Rey, S.~Theisen and J.~T.~Yee,
  Nucl.\ Phys.\ B {\bf 527}, 171 (1998)
  [hep-th/9803135].
  

\bibitem{Kharzeev:2014pha} 
  D.~E.~Kharzeev,
  Phys.\ Rev.\ D {\bf 90}, no. 7, 074007 (2014)
  [arXiv:1409.2496 [hep-ph]].

\bibitem{Hashimoto:2014fha} 
  K.~Hashimoto and D.~E.~Kharzeev,
  Phys.\ Rev.\ D {\bf 90}, no. 12, 125012 (2014)
  [arXiv:1411.0618 [hep-th]].

\bibitem{Iatrakis:2015sua} 
  I.~Iatrakis and D.~E.~Kharzeev,
  Phys.\ Rev.\ D {\bf 93}, no. 8, 086009 (2016)
  doi:10.1103/PhysRevD.93.086009
  [arXiv:1509.08286 [hep-ph]].

\bibitem{Fadafan:2015ynz} 
  K.~Bitaghsir Fadafan and S.~K.~Tabatabaei,
  Phys.\ Rev.\ D {\bf 94}, no. 2, 026007 (2016)
  doi:10.1103/PhysRevD.94.026007
  [arXiv:1512.08254 [hep-ph]].

\bibitem{Iida:2013qwa} 
  H.~Iida, T.~Kunihiro, B.~Mueller, A.~Ohnishi, A.~Schaefer and T.~T.~Takahashi,
  Phys.\ Rev.\ D {\bf 88}, 094006 (2013)
  doi:10.1103/PhysRevD.88.094006
  [arXiv:1304.1807 [hep-ph]].

\bibitem{Tsukiji:2016krj} 
  H.~Tsukiji, H.~Iida, T.~Kunihiro, A.~Ohnishi and T.~T.~Takahashi,
  Phys.\ Rev.\ D {\bf 94}, no. 9, 091502 (2016)
  doi:10.1103/PhysRevD.94.091502
  [arXiv:1603.04622 [hep-ph]].

\bibitem{Tsukiji:2017pjx} 
  H.~Tsukiji, T.~Kunihiro, A.~Ohnishi and T.~T.~Takahashi,
  PTEP {\bf 2018}, no. 1, 013D02 (2018)
  doi:10.1093/ptep/ptx186
  [arXiv:1709.00979 [hep-ph]].

\bibitem{Fukushima:2013dma} 
  K.~Fukushima,
  Phys.\ Rev.\ C {\bf 89}, no. 2, 024907 (2014)
  doi:10.1103/PhysRevC.89.024907
  [arXiv:1307.1046 [hep-ph]].

\bibitem{Fukushima:2016xgg} 
  K.~Fukushima,
  Rept.\ Prog.\ Phys.\  {\bf 80}, no. 2, 022301 (2017)
  doi:10.1088/1361-6633/80/2/022301
  [arXiv:1603.02340 [nucl-th]].
    
\bibitem{Hashimoto:2014xta} 
  K.~Hashimoto, S.~Kinoshita, K.~Murata and T.~Oka,
  Phys.\ Lett.\ B {\bf 746}, 311 (2015)
  doi:10.1016/j.physletb.2015.05.004
  [arXiv:1408.6293 [hep-th]].
  
\bibitem{Hashimoto:2014dda} 
  K.~Hashimoto, S.~Kinoshita, K.~Murata and T.~Oka,
  Nucl.\ Phys.\ B {\bf 896}, 738 (2015)
  doi:10.1016/j.nuclphysb.2015.05.004
  [arXiv:1412.4964 [hep-th]].
  
\bibitem{Hashimoto:2015psa} 
  K.~Hashimoto, M.~Nishida and A.~Sonoda,
  JHEP {\bf 1508}, 135 (2015)
  doi:10.1007/JHEP08(2015)135
  [arXiv:1504.07836 [hep-th]].
  
\bibitem{Hashimoto:2016wme} 
  K.~Hashimoto, K.~Murata and K.~Yoshida,
  Phys.\ Rev.\ Lett.\  {\bf 117}, no. 23, 231602 (2016)
  doi:10.1103/PhysRevLett.117.231602
  [arXiv:1605.08124 [hep-th]].

\bibitem{Arias:2009me} 
  R.~E.~Arias and G.~A.~Silva,
  JHEP {\bf 1001}, 023 (2010)
  [arXiv:0911.0662 [hep-th]].



\bibitem{Schmidhuber:1996fy} 
  C.~Schmidhuber,
  Nucl.\ Phys.\ B {\bf 467}, 146 (1996)
  [hep-th/9601003].

\bibitem{Cai:2000yk} 
  R.~G.~Cai and N.~Ohta,
  Prog.\ Theor.\ Phys.\  {\bf 104}, 1073 (2000)
  [hep-th/0007106].





\end{thebibliography}
\end{document}